\documentclass[epsfig,12pt]{article}
\usepackage{amssymb}
\usepackage{graphicx,amsmath}

\begin{document}

\newcommand \be  {\begin{equation}}
\newcommand \bea {\begin{eqnarray} \nonumber }
\newcommand \ee  {\end{equation}}
\newcommand \eea {\end{eqnarray}}
\def\x{\mathbf{x}}

\title{ High-Dimensional Random Fields and Random Matrix Theory\footnote{Dedicated to Prof. Leonid Pastur on the occassion of his 75th birthday}}

\author{Yan V. Fyodorov\\{\small School of Mathematical Sciences, Queen Mary University of London,}\\ {\small London E1 4NS, United Kingdom}}

\date{November 18, 2013}

\maketitle

\begin{abstract}
     Our goal is to discuss in detail the calculation of the mean number of stationary points and minima for random isotropic Gaussian fields on a sphere as well as for stationary Gaussian random fields in a background parabolic confinement. After developing the general formalism based on the high-dimensional Kac-Rice formulae we combine it with the Random Matrix Theory (RMT) techniques to perform  analysis of the random energy landscape of $p-$spin spherical spinglasses and a related glass model, both displaying a zero-temperature one-step replica symmetry breaking glass transition as a function of control parameters (e.g. a magnetic field or curvature of the confining potential). A particular emphasis of the presented analysis is on understanding in detail the picture of "topology trivialization" (in the sense of drastic reduction of the number of stationary points) of the landscape which takes place in the vicinity of the zero-temperature glass transition in both models. We will reveal the important role of the GOE "edge scaling" spectral region and the Tracy-Widom distribution of the maximal eigenvalue of GOE matrices for providing an accurate quantitative description of the universal features of the topology trivialization scenario.
 \end{abstract}

\section{Introduction}

Understanding statistical structure of stationary points (minima, maxima and saddles) of random landscapes and fields of various types and dimensions is a rich problem of intrinsic current interest in diverse areas of pure and applied mathematics and mathematical physics~\cite{math1,math2,LGH,Baugher,Auf1,Auf2,Nicolaescu1,Nicolaescu2,Nicolaescu3,FZ}. In particular, most recently it was understood that methods and techniques  used to study this type of questions prove to be useful for characterization of generic topological properties of real algebraic varieties \cite{GW1,GW2,FLL}. On the other hand, the same problem keeps attracting steady interest in theoretical physics community over more than fifty years with a range of applications in condensed matter theory \cite{halp1}, classical and quantum optics \cite{halp2,Freund1995,KleinAgam}, physics of  glasses and spin glasses \cite{Lennard-Jones,CGP,Fyo04,my2005,spinglass1,spinglass2,spinglass2a,BD07,FyoWi07,FyNa12,MSK}, string theory~\cite{string1,string2} and cosmology~\cite{cosm1,cosm2,cosm3}. The above list is hopefully representative, but surely not exhaustive.

The goal of the present notes, written in an informal style of theoretical physics, is to discuss in detail the calculation of the mean total number of stationary points and the minima on the example of a simple, but rich model of general interest: random isotropic Gaussian fields on a sphere of any dimension $N$. We also briefly discuss in the end the case of stationary (i.e. statistically translationally-invariant) random fields plus a deterministic confining parabolic potential.  Starting from the general Kac-Rice formulae (\ref{KR}) and  (\ref{KRMin}) we will gradually arrive to our central result encapsulated in two expressions
 (\ref{GOE7})-(\ref{GOE7prim}) and (\ref{min7})-(\ref{min8}) which provide explicit evaluations of the mean quantities in question by relating them to statistics of the eigenvalues of random matrices from the Gaussian Orthogonal Ensemble. We will then proceed to asymptotic analysis of a representative high-dimensional $N\gg 1$ variant of the model: the so-called $p-$spin spherical spinglass. Our particular emphasis will be on understanding the critical behaviour of the mean number of stationary points and the minima in the small vicinity of a zero-temperature spin-glass transition which occurs in such a model with growing external magnetic field. Such transition can be understood as gradual "topology trivialization" of the landscape, in the sense introduced in \cite{FLD2013}, that is a drastic reduction of the number of stationary points which takes place accross the transition region. We will reveal the important role of the GOE "edge scaling" spectral region and the Tracy-Widom distribution of the maximal eigenvalue of GOE matrices for providing an accurate quantitative description of the topology trivialization. In the last section we give a short account on related issues for stationary isotropic random fields, and then briefly discuss related and outstanding questions which would be interesting to address.

Before indulging into calculations a few remarks on the history of the subject are due. Presence of the modulus of the Hessian in the Kac-Rice formula (\ref{KR}) or the indicator function in (\ref{KRMin}) for a long time was considered to be a serious
obstacle for providing general evaluation of the mean number of stationary points, see e.g. \cite{Kurchan}. It seems that the idea of putting the problem into the Random Matrix Theory context by relating the mean counting functions to the mean density of  GOE eigenvalues has appeared originally in\cite{Fyo04}, and has been further developed in \cite{my2005,BD07,FyoWi07}.  It was then independently rediscovered in \cite{Auf1}. The latter paper (and its sequel \cite{Auf2}) considerably advanced that technique and provided a few important insights into counting stationary points for random isotropic Gaussian fields on a sphere, conditioned to have a fixed index and a fixed value of the field at the stationary point. Attempts of the present author to understand better some parts of calculations presented in \cite{Auf1,Auf2}, as well as his own work  \cite{FyoWi07,FyNa12} on the number of minima for translationally-invariant random fields without spherical constraints provided both the background and an incentive for writing the present notes. The method of arriving to the results in \cite{Auf1,Auf2} was essentially relying on the differential geometry approach to the problem developed earlier in \cite{math1}. Unfortunately two elegant formulae (\ref{GOE7}) and (\ref{min7}) which proved to be very useful for the analysis of  landscape topology of spherical spin-glass model, as well as for applications to statistical topology of real algebraic varieties \cite{FLL} have not appeared in \cite{Auf1,Auf2} in an explicit form. In fact, they were discovered rather recently in the course of applying the results of \cite{Auf1,Auf2} to the analysis of the topology trivialization phenomenon in the simplest $p=2$ case of the spherical spin glass, see \cite{FLD2013} for more detail.
 The author believes that an ab initio derivation of (\ref{GOE7}) and (\ref{min7}) explained in considerable detail in Sec.2.1-2.2 and Sec.3 of the notes without any recourse to the results of \cite{math1} may have some methodological merits on its own. Sec. 2.3.2 is closely related to the calculation which appeared originally in \cite{FLD2013}, and Sec. 4 is basically a review of author's earlier works \cite{Fyo04} and \cite{FyNa12}. Beyond that, substantial parts of the analysis presented in Sec. 2.3 and Sec. 3 of the article, as well as the content of the Appendix are probably new.

\subsubsection*{Acknowledgements}
The text of this article is a modified and extended version of the lecture notes for the Summer School " Randomness in Physics and Mathematics" which took place at ZIF, Bielefeld, August 5-17, 2013. The author
is grateful to the organizers of the school for kind hospitality and to participants for an informative feedback.  Some results discussed here first appeared in works of the author with C. Nadal and P. Le Doussal whose collaboration is acknowledged with gratitude. The author is also grateful to Antonio Lerario for explaining relevance of the exposed methods and results to problems in topology of real algebraic varieties, and to Dan Cheng for a stimulating question about stationary anisotropic random fields which resulted in adding the Appendix. This research was supported by EPSRC grant EP/J002763/1 ``Insights into Disordered Landscapes via Random Matrix Theory and Statistical Mechanics''.

\subsection{Kac-Rice formulae in one and higher dimensions}

 Let $v(x), \, x\in \mathbb{R}$ be any at least once differentiable real valued function and let $\delta(x)$ stand for the Dirac $\delta-$function which we may informally think of as defined by its Fourier-integral representation: $ \delta(x)=\lim_{\epsilon\to 0^{+}} \delta_{\epsilon}(x)$ where $\delta_{\epsilon>0}(x)=\frac{1}{2\pi}\int_{-\infty}^{\infty}e^{ikx-\epsilon \frac{k^2}{2}}\,dk\,$. Then the integral
\begin{equation}\label{Gauproc3}
N_V(a,b)=\int_{a}^{b}\delta(v(x)-V)\left|\frac{dv}{dx}\right|\,dx
\end{equation}
 yields the number of crossings of the level $V$ by the curve $v(x)$  in the interval $(a,b)$, i.e. the number of real roots of the equation
 $v(x)=V$ in that interval, tacitly assuming that each root is simple. This formula is nothing else
 but an incarnation of the fundamental property of the Dirac delta-function:
 \begin{equation}\label{dir9}
 \delta\left[f(x))\right]=\sum_{x_n:\{f(x_n)=0\}}\frac{1}{|f'(x_n)|}\delta(x-x_n).
 \end{equation}

If the function $v(x)$ is random, the number of crossings/roots is also random and one can be interested in its statistical characteristics like moments. In what follows we are going to consider only Gaussian random functions
such that the joint probability density (j.p.d) of $v_1=v(x_1),v_2=v(x_2)\ldots,v_k=v(x_k)$
 for any choice of $x_1,x_2,\ldots, x_k$ is the density of a multivariate Gaussian distribution. The definition can be trivially extended to Gaussian functions of many variables, or fields.
  The Gaussian processes are characterized by the mean $\mathbb{E}\left\{v(x)\right\}$ and the covariance $\mathbb{E}\left\{v(x_1)v(x_2)\right\}$, with  $\mathbb{E}\left\{A\right\}$ denoting here and henceforth the expected (or, equivalently, the mean) value of a random variable $A$. If the mean of a process is independent of $x$, and the covariance depends only on the difference $x_1-x_2$ the process is called stationary.

  The simplest nontrivial information is given by the mean number of crossings $\mathbb{E}\{N_V(a,b)\}=\int_{a}^{b}\mathbb{E}\left\{\delta(v(x)-V)\left|\frac{dv}{dx}\right|\right\}\,dx$.
Expressions of this type are traditionally called in the literature {\it Kac-Rice formulae} and go back to seminal works \cite{Rice,Kac1,Kac2}. The same approach can be also straightforwardly adopted to finding the mean value of {\it extrema}
(minima and maxima) of smooth enough (i.e. at least twice-differentiable) function $f(x)$, as such extrema
correspond to the roots of the equation $f'(x)=0$, hence their mean number in the interval $[a,b]$ is given by the Kac-Rice formula
\begin{equation}\label{extr}
{\cal N}(a,b)=\int_{a}^{b}\mathbb{E}\left\{\delta\left(f'(x)\right)\left|f''(x)\right|\right\}\,dx,
\end{equation}
where the expectation is over the joint probability density of the derivatives $f'(x),f''(x)$ at the same point $x$\footnote{In the mathematical literature the Kac-Rice type formulae like (\ref{extr}) and their multidimensional analogues like (\ref{KR}) are most frequently written in the form: ${\cal N}(a,b)=\int_{a}^{b}\mathbb{E}\left\{|f''(x)|\,\,{\bf \|}\,\,  f'(x)=0\right\}\,dx$ where $\mathbb{E}\left\{{\bf var}\,\, {\|}\,\, {\bf cons}\right\}$ stands for the conditional expectation of the variable ${\bf var}$ given the constraints ${\bf cons}$. We will however keep using the Dirac delta-function as is common in the physical literature.}. Actually, it is possible to represent the mean number of minima alone by a similar formula. Namely, introducing the Heaviside indicator function $\theta(A)=1$ if $A>0$ and zero otherwise, we can write
\begin{equation}\label{extr1}
{\cal N}_{min}(a,b)=\int_{a}^{b}\mathbb{E}\left\{\delta\left(f'(x)\right)f''(x)\,\theta(f''(x))\right\}\,dx\,.
\end{equation}
Note that had we omitted the indicator function in the above formula (or equivalently suppressed the modulus in  (\ref{extr})) the value of the integral would yield the difference between the number of minima and the number of
maxima of the function, which is the topological invariant depending only on the boundary values of $f(x)$ for $x\to \pm \infty$.

It is natural to ask similar questions about extrema of random fields, i.e. random functions of several variables, which we will frequently also call {\sf random landscapes} anticipating their use in Statistical Mechanics context.
Assuming that a landscape is described by a sufficiently smooth random function $V({\bf x})$
of $N$ real variables ${\bf x}=(x_1,...,x_N)$ we will be in general interested in
counting the expected number of its stationary points of different
types: minima, maxima and saddles of various indices. The
simplest, yet already a non-trivial problem of this sort is to
find the mean number $\mathbb{E}\left\{{\cal N}_s\right\}$ of {\it all} stationary
points, irrespective of their index.  The problem amounts
to finding all solutions of the simultaneous stationarity
conditions $\partial_k  V=0$ for all $k=1,...,N$, with
$\partial_k$ standing for the partial derivative
$\frac{\partial}{\partial x_k}$. The total number ${\cal N}_s(D)$ of the
stationary points in any spatial domain $D\in \mathbb{R}^N$ is then given by
${\cal N}_s(D)=\int_D \rho_s({\bf x}) \, d{\bf x}$, with $\rho_s({\bf
x})$ being the corresponding density of the stationary points. The
mean value of such a density can be found according
to the multidimensional analogue of the Kac-Rice
formula:
\begin{equation}\label{KR}
\mathbb{E}\left\{\rho_{s}({\bf x})\right\}=\mathbb{E}\left\{
|\det{\left(\partial^2_{k_1,k_2} V\right)|
\prod_{k=1}^N\delta(\partial_k V)}\right\},
\end{equation}
see \cite{math1,math2,math3} for various rigorous derivations.
Note again the importance of keeping the modulus of the determinant of the Hessian matrix $\partial^2_{k_1,k_2} V$ in (\ref{KR}), as omitting it would yield instead the density of
the object related to the Euler characteristics of the surface which is the main subject of
 the book \cite{math1}.

Similarly, if one is interested in counting only minima, the
corresponding mean density can be written as
\begin{equation}\label{KRMin}
\mathbb{E}\left\{\rho_{m}({\bf x})\right\}=\mathbb{E}\left\{
\det{\left(\partial^2_{k_1,k_2} V\right)}\theta\left(\partial^2_{k_1,k_2} V\right)
\prod_{k=1}^N\delta(\partial_k V)\right\},
\end{equation}
where here and henceforth the value of the indicator function of matrix argument $\theta\left(K\right)$ is chosen to be unity if a real symmetric matrix $K$ is positive definite and zero if at least one of the eigenvalues of $K$ is negative. Such choice
selects only stationary points with positive definite Hessians,
which are minima.

The main goal of the present notes is to provide a detailed analysis of the above Kac-Rice formulae for the case of isotropic high-dimensional random Gaussian fields on spheres of any dimension.  This will help us to get insights into the properties of random energy landscapes of one of the simplest, yet paradigmatic  models used in physics of disordered systems, the so-called spherical spin glass. We shall see that the methods and results of the theory of random matrices, standardly abbreviated as RMT, play central role in that study. We will also review the analogous results for Gaussian stationary fields in full euclidean space, and apply it to counting stationary points and minima for another type of high-dimensional random landscapes. This will allow us to reveal universal and non-universal feature of this type of problems.

\section{The mean number of stationary points for isotropic Gaussian random fields on a sphere.}

\subsection{General formalism.}
A (mean-zero) Gaussian random field (or landscape) $V(x_1,\ldots,x_N)$ defined in $N-$dimensional Euclidean space $\x \in \mathbb{R}^N$  is called {\sf isotropic} if its covariance
$\mathbb{E}\left\{V(\x)V(\x')\right\}$  is invariant with respect to a simultaneous rotation of the two vectors: $\x\to O\x,\, \x'\to O\x' $,  with $O\in O(N)$ being any $N\times N$ orthogonal matrix. This implies
that the covariance may depend only on the scalar product $\x\cdot \x'=\sum_{i=1}^Nx_ix'_i$ of the vectors $\x$ and $\x'$ as well as on their lengths $|\x|,|\x'|$. For the purposes of these lectures it is enough to consider the dependence on the first argument only, so that we assume
\begin{equation}\label{coviso}
\mathbb{E}\left\{V(\x)V(\x')\right\}=F(\x\cdot \x').
\end{equation}
Denote by ${\cal V}$ the restriction of such a field to the surface of a sphere $S_{N-1}(R)$ of radius $R$ defined by $S_{N-1}(R):\{x_1^2+\ldots + x_N^2=R^2\}$. It will be convenient to think of the sphere as a union of the northern $x_N\ge 0$ and the southern $x_N\le 0$
 hemispheres, and since the statistics of stationary points of ${\cal V}$ must be obviously identical in the two hemispheres due to the rotational invariance of the field we can restrict our attention to properties of  ${\cal V}$ in one of them, say the northern one.
There we can look at ${\cal V}$ as a random function of $N-1$ variables $x_1,\ldots, x_{N-1}$
defined in the domain $D: \{\sum_{j=1}^{N-1}x_j^2\le R^2 \}$ via the relation:
\begin{equation}\label{den1}
{\cal V}(x_1,\ldots,x_{N-1})=V\left(x_1,\ldots, x_{N-1}, x_N=\sqrt{R^2-\sum_{j=1}^{N-1}x_j^2} \right).
\end{equation}
In what follows we find it convenient to use for brevity the $N-1$ component vector notation $\tilde{\x}=(x_1,\ldots,x_{N-1})$ and
$\tilde{\x}^T$ for the transposed vector.
 For the later use it is expedient to write down the Euclidean length element (the first fundamental form)
 of the hemisphere as
 \begin{equation}\label{Riemannian}
 (dl)^2= \sum_{j=1}^{N}(dx_j)^2=\sum_{l,m}^{N-1}g_{lm}(\tilde{\x})dx_ldx_m, \quad g_{lm}(\tilde{\x})=\delta_{lm}+\frac{x_lx_m}{R^2-\tilde{\x}^2},
\end{equation}
where the matrix ${\bf g}={\bf 1}+\frac{1}{R^2-\tilde{\x}^2}\tilde{\x}\otimes \tilde{\x}^T$ of coefficients $g_{lm}(\x)$ defines the Riemannian metric on the hemisphere. We will also need the inverse and the determinant of that matrix:
\begin{equation}\label{invdet}
{\bf g}^{-1}={\bf 1}-\frac{1}{R^2}\tilde{\x} \otimes \tilde{\x}^T, \quad \det{\bf g}=\frac{R^2}{R^2-\tilde{\x}^2}.
\end{equation}
 In particular, according to the general principles of the Riemannian geometry the volume (surface area) of the hemisphere can be calculated as
\begin{equation}\label{Rvolume}
V_{N-1}(R)=\int_{D}\,\sqrt{\det{\bf g}}\, dx_1dx_2 \cdots dx_{N-1}
\end{equation}
\begin{equation}\label{Rvolume1}
= \int_{\tilde{\x}^2\le R^2}\,\,\frac{R}{\sqrt{R^2-\tilde{\x}^2}}\, \prod_{j=1}^{N-1} dx_j=R^{N-1}\frac{\pi^{N/2}}{\Gamma(N/2)},
\end{equation}
where $\Gamma(z)$ is the Euler's gamma-function.

According to the general Kac-Rice formula (\ref{KR}) the mean number $\mathbb{E}\left\{{\cal N}^{+}_s\right\}$ of all stationary points of the random field ${\cal V}(\tilde{\x})$ belonging to the northern hemisphere is given by
\begin{equation}\label{KRsph1}
\mathbb{E}\left\{{\cal N}^{+}_s\right\}=\int_D \mathbb{E}\left\{|\det{\left(\partial^2_{k_1,k_2} {\cal V}\right)|
\prod_{k=1}^{N-1}\delta(\partial_k {\cal V})}\right\}\,\prod_{j=1}^{N-1} dx_j.
\end{equation}
The first step towards evaluating this integral amounts to calculating the joint probability density of $N-1$ first derivatives $ \partial_k {\cal V}$ and $N(N-1)/2$ second derivatives $\partial^2_{k_1,k_2} {\cal V}$ taken at the same spatial point $\tilde{\x}$. As those are mean-zero Gaussian variables, the task in turn amounts to calculating the corresponding covariances. This can be straightforwardly done using the covariance structure of the field ${\cal V}$ inherited from (\ref{coviso}), that is:
\begin{equation}\label{KRsph3}
\mathbb{E}\left\{{\cal V}(\tilde{\x}){\cal V}(\tilde{\x}')\right\}=F\left(\tilde{\x}\cdot\tilde{\x}'+\sqrt{R^2-\tilde{\x}^2}\sqrt{R^2-\tilde{\x}'^2}\right).
\end{equation}
For example, we have:
\[
\mathbb{E}\left\{\partial_i {\cal V}(\tilde{\x})\partial_j {\cal V}(\tilde{\x})\right\}
\]
\begin{equation}\label{KRsph4}
=\frac{\partial}{\partial x_i}
\frac{\partial}{\partial x'_j}
F\left(\tilde{\x}\cdot\tilde{\x}'+\sqrt{R^2-\tilde{\x}^2}\sqrt{R^2-\tilde{\x}'^2}\right)|_{\x=\x'}=g_{ij}(\tilde{\x})F'(R^2).
\end{equation}
In a similar manner a straightforward but somewhat lengthy differentiation yields also the covariances:
\begin{equation}\label{KRsph5}
\mathbb{E}\left\{\partial^2_{ik} {\cal V}(\tilde{\x})\partial_j {\cal V}(\tilde{\x})\right\}
=g_{ik}(\tilde{\x})\frac{x_j}{R^2-\tilde{\x}^2}F'(R^2)
\end{equation}
and finally
\[
\mathbb{E}\left\{\partial^2_{ik} {\cal V}(\tilde{\x})\partial^2_{jl} {\cal V}(\tilde{\x})\right\}=g_{ik}(\tilde{\x})g_{jl}(\tilde{\x})\frac{1}{R^2-\tilde{\x}^2}F'(R^2)
\]
\begin{equation}\label{KRsph6}
+\left\{g_{ij}(\tilde{\x})g_{kl}(\tilde{\x})+
g_{jk}(\tilde{\x})g_{il}(\tilde{\x})+g_{ik}(\tilde{\x})g_{jl}(\tilde{\x})\right\}F''(R^2).
\end{equation}
In particular, the formula (\ref{KRsph5}) shows that the vector ${\bf v}=(\partial_1 {\cal V}, \ldots,\partial_{N-1} {\cal V}) $ of the first derivatives and the Hessian matrix $K_{ik}=\partial^2_{ik} {\cal V}$ of second derivatives are not independent as their mutual covariance is nonvanishing.  This fact, which distinguishes isotropic Gaussian fields on the sphere from their stationary counterparts in the full Eucliden space (see the Appendix for the latter case) makes the problem of evaluating the expectation in the equation (\ref{KRsph1}) looking somewhat problematic.

To cope with the problem one may introduce a new random matrix $\kappa_{ik}$ whose elements are
related to $K_{ik}$ by a shift linear in the variables ${\bf v}$, namely
\begin{equation} \label{KRsph7}
\kappa_{ik}=K_{ik}-\frac{g_{ik}(\tilde{\x})}{R^2-\tilde{\x}^2}\, \left({\bf v}^T{\bf g}^{-1}\,\tilde{\x}\right),
\end{equation}
where ${\bf g}^{-1}$ stands for the inverse of the matrix ${\bf g}(\tilde{\x})$, see (\ref{invdet}).
Then straightforward calculations exploiting (\ref{KRsph4}) and (\ref{KRsph5}) show that the mutual covariances $\mathbb{E}\{\kappa_{ik}v_m\}$ vanish identically for any choice of the indices $i,k$ and $m$,  hence the corresponding Gaussian variables are independent. This can be readily exploited in the integral (\ref{KRsph1}).
 Namely, we can simply replace random variables  $\partial^2_{ik} {\cal V}\equiv K_{ik}$ in the integrand with $\kappa_{ik}$ since the difference between the two sets of the variables is proportional to components $\partial_{i} {\cal V}$ of the vector ${\bf v}$ and vanishes due to $\delta-$function factors in the integrand. After doing this
the expectation in (\ref{KRsph1}) decouples into the product of two factors, and since the used change of variables yields the trivial Jacobian equal to unity we can bring
the formula (\ref{KRsph1}) to the form:
\begin{equation}\label{KRsph8}
\mathbb{E}\left\{{\cal N}_s\right\}=\int_D \mathbb{E}\left\{|\det{\left(\kappa_{ik}\right)}|\right\}
\mathbb{E}\left\{\prod_{k=1}^{N-1}\delta(v_k)\right\}\,\prod_{j=1}^{N-1} dx_j.
\end{equation}
To calculate the second factor in the integrand of (\ref{KRsph8}) is easy as the j.p.d. ${\cal P}({\bf v})$ of the components of the vector ${\bf v}$ according to (\ref{KRsph4}) is given by
\begin{equation}\label{KRsph9}
{\cal P}({\bf v})=\frac{1}{[2\pi F'(R^2)]^{(N-1)/2}\sqrt{\det{{\bf g}(\tilde{\x})}}} \exp{\left\{-\frac{1}{2F'(R^2)} \left({\bf v}^T{\bf g}^{-1}\,{\bf v} \right)\right\}}
\end{equation}
which immediately yields
\begin{equation}\label{KRsph10}
\mathbb{E}\left\{\prod_{k=1}^{N-1}\delta(v_k)\right\}=\frac{1}{[2\pi F'(R^2)]^{(N-1)/2}\sqrt{\det{{\bf g}(\tilde{\x})}}}.
\end{equation}

To evaluate the first factor in the integrand of (\ref{KRsph8}) one first needs to determine the covariance structure of the variables $\kappa_{ik}$ defined in (\ref{KRsph7}) by exploiting (\ref{KRsph5}) and (\ref{KRsph6}).
 After straightforward but somewhat lengthy calculations one arrives at
 \[
 \mathbb{E}\{\kappa_{ik}\kappa_{jl}\}=\frac{F'(R^2)}{R^2}g_{ik}(\tilde{\x})g_{jl}(\tilde{\x})
 \]
 \begin{equation}\label{KRsph11}
+\left\{g_{ij}(\tilde{\x})g_{kl}(\tilde{\x})+
g_{jk}(\tilde{\x})g_{il}(\tilde{\x})+g_{ik}(\tilde{\x})g_{jl}(\tilde{\x})\right\}F''(R^2)
\end{equation}
 Finally, one can further notice that it is convenient to  pass from $\kappa_{ik}$ to a new set of variables $\tilde{\kappa}_{ik}$ by introducing the matrix $\tilde{\kappa}={\bf g}^{-1/2} \kappa {\bf g}^{-1/2}$.
  The point of such a change is that the covariance structure of the variables $\tilde{\kappa}_{ij}$ turns out to be independent of the metric ${\bf g}(\tilde{\x})$,
hence totally coordinate-independent:
 \begin{equation}\label{KRsph12}
 \mathbb{E}\{\tilde{\kappa}_{ik}\tilde{\kappa}_{jl}\}=\frac{F'(R^2)}{R^2}\delta_{ik}\delta_{jl}+F''(R^2)
 \left\{\delta_{ij}\delta_{kl}+\delta_{jk}\delta_{il}+\delta_{ik}\delta_{jl}\right\}
\end{equation}

Denoting the joint probability density of such coordinate-independent random matrix $\tilde{\kappa}$ as ${\cal P}_0(\tilde{\kappa})$,  using $|\det{\kappa}|=\det{{\bf g}}\cdot |\det{\tilde{\kappa}}|$ and denoting $d{\tilde{\kappa}}=\prod_{i\le j}d \tilde{\kappa}_{ij} $ we have
\begin{equation}\label{KRsph13}
 \mathbb{E}\left\{|\det{\left(\kappa\right)}|\right\}=\det{{\bf g}}\cdot \mathbb{E}\left\{|\det{\left(\tilde{\kappa}\right)}|\right\}=\det{{\bf g}}\,\int {\cal P}_0(\tilde{\kappa} ) |\det{\tilde{\kappa}}|\, d{\tilde{\kappa}}\,.
 \end{equation}
 Now we can substitute (\ref{KRsph13}) and (\ref{KRsph10}) into (\ref{KRsph8}) and notice that the coordinate-dependent factors conspired precisely to yield upon the integration the surface area $V_{N-1}(R)$ of the hemisphere given by (\ref{Rvolume1}). Multiplying this result by the factor of two we arrive to the expression for the mean number of all stationary points of an isotropic random field on the whole sphere given by
 \begin{equation}\label{KRsph14}
\mathbb{E}\left\{{\cal N}_s\right\}=2V_{N-1}(R)\,\frac{1}{[2\pi F'(R^2)]^{(N-1)/2}}\,\int {\cal P}_0(\tilde{\kappa} ) |\det{\tilde{\kappa}}|\, d{\tilde{\kappa}}\,.
\end{equation}
To proceed further we need to find an explicit expression for the probability density ${\cal P}_0(\tilde{H} )$
of $n\times n$ random matrix $H=\tilde{\kappa}$  corresponding to the covariance structure (\ref{KRsph12})
with the final choice $n=N-1$. To that end, let $K,H$ stand for real symmetric $n\times n$ matrices, and $H=\tilde{\kappa}$  with the mean zero Gaussian-distributed entries $\tilde{\kappa}_{ij}$ characterized by the covariance structure (\ref{KRsph12}). Using the identity
\begin{equation}\label{KRsph15}
{\cal P}_0(H)=\mathbb{E}\left\{\delta(H-\tilde{\kappa} )\right\}=\frac{1}{2^n (2\pi)^{n(n+1)/2}}\int e^{\frac{i}{2}{\small \mbox{Tr}}(KH)}
\mathbb{E}\left\{e^{-\frac{i}{2}{\small \mbox{Tr}}(K\tilde{\kappa})}\right\}\, dK
\end{equation}
it is straightforward to show that
\begin{equation}\label{KRsph16}
{\cal P}_0(H)=\frac{1}{2^{n/2}}(\pi a)^{-n(n+1)/4}\int_{-\infty}^{\infty} e^{-\frac{1}{2a}{\small \mbox{Tr}
\left(t\sqrt{b}\, {\bf 1}-H\right)^2}}\, e^{-\frac{t^2}{2}} \frac{dt}{\sqrt{2\pi}}
\end{equation}
where we have denoted
\begin{equation}\label{KRsph16a}
a=2F''(R^2), \quad b=\frac{1}{R^2}F'(R^2)+F''(R^2)
\end{equation}

Exploiting the above result and remembering $n=N-1$ we therefore can represent (\ref{KRsph14}) in the form
 \begin{equation}\label{KRsph17}
\mathbb{E}\left\{{\cal N}_s\right\}=2\frac{V_{N-1}(R)}{2^{(N-1)/2}}\,\frac{1}{[2\pi F'(R^2)]^{(N-1)/2}}\,\frac{1}{[2\pi F''(R^2)]^{N(N-1)/4}}
\end{equation}
\[
\times \int_{-\infty}^{\infty}e^{-\frac{t^2}{2}} \frac{dt}{\sqrt{2\pi}}\,\int |\det{H}|e^{-\frac{1}{2a} \mbox{\small Tr}\left(t\sqrt{b}\, {\bf 1}-H\right)^2}\, \, dH
\]
After further changing the integration variable $t\to t/\sqrt{b}$ and then replacing $H\to t{\bf 1}-H$ the above expression assumes a form most convenient for further analysis:
 \begin{equation}\label{KRsph17}
\mathbb{E}\left\{{\cal N}_s\right\}=2\frac{V_{N-1}(R)}{2^{(N-1)/2}\sqrt{b}}\,\frac{1}{[2\pi F'(R^2)]^{(N-1)/2}}\,\frac{1}{[2\pi F''(R^2)]^{N(N-1)/4}}
\end{equation}
\[
\times \int_{-\infty}^{\infty}e^{-\frac{t^2}{2b}} \frac{dt}{\sqrt{2\pi}}\,\int \, e^{-\frac{1}{2a}{\small \mbox{Tr}H^2}} \,|\det{(t{\bf 1}-H)}|\, \, dH\,,
\]
where the last integral goes over $(N-1)\times (N-1)$ real symmetric matrices.
Further progress requires the use of ideas from the Random Matrix Theory (RMT), with the books by Forrester\cite{Forrbook} or Mehta\cite{Mehtabook} being useful introductions to the subject.

\subsection{Relation to the Gaussian Orthogonal Ensemble}
The Gaussian Orthogonal Ensemble (GOE) is defined as the probability measure on the space of real symmetric
$n\times n$ matrices $H$:
 \begin{equation}\label{GOE1}
d\mu^{(GOE)}_{n,a}(H)=C_{n}(a)e^{-\frac{1}{2a}{\small \mbox{Tr}H^2}} \, dH, \quad a>0
\end{equation}
where $C_{n}(a)=(\pi a)^{-\frac{n(n+1)}{4}}2^{-n/2}$ is the appropriate normalization constant ensuring $\int d\mu^{(GOE)}_{n,a}(H)=1$.
This definition is equivalent to requiring all entries $H_{i\le j}$ to be independent mean zero real Gaussian variables with variances:  $\mathbb{E}\left\{H^2_{ii}\right\}=a, \, \mathbb{E}\left\{H^2_{i<j}\right\}=a/2$.
The measure is invariant with respect to conjugating $H\to OHO^{-1}$, with orthogonal $n\times n$ matrices $O=O^T$ forming the Orthogonal Group $O(n)$, hence the name of the ensemble. Recall that any real symmetric matrix $H$
can be diagonalized by an orthogonal conjugation: $H=O^{-1}\Lambda O,\,\,O\in O(n)$, where
$\Lambda=\mbox{diag}(\lambda_1,\ldots,\lambda_n)$ is the diagonal matrix of real eigenvalues.
 This implies that for any invariant function such that $\phi(H)=\phi(OHO^{-1}), \forall O\in O(n)$ we can write:
  \begin{equation}\label{GOE2}
\mathbb{E}\left\{\phi(H)\right\}=\int \phi(H)\,d\mu^{(GOE)}_{n,a}(H)=Z^{-1}_{n}(a)\int_{\mathbb{R}^N} \phi(\Lambda) e^{-\frac{1}{2a}\sum_{i=1}^n \lambda_i^2} |\Delta_n(\Lambda)|\, d\Lambda, \quad
\end{equation}
where the Jacobian factor $\Delta_n(\Lambda)=\prod_{i<j}(\lambda_i-\lambda_j)$ is the so-called Vandermonde determinant. The associated normalization constant $Z_{n}(a)$ is given by a particular instance of the so-called Selberg integral \cite{Forrbook,Mehtabook}:
 \begin{equation}\label{GOE3}
Z_{n}(a)=\int_{\mathbb{R}^N} \,e^{-\frac{1}{2a}\sum_{i=1}^n \lambda_i^2} |\Delta_n(\Lambda)|\, d\Lambda  =(2\pi)^{n/2}a^{n(n+1)/4}\prod_{j=1}^n\frac{\Gamma\left(1+\frac{j}{2}\right)}{\Gamma\left(\frac{3}{2}\right)}.
\end{equation}
 The simplest spectral characteristic of the GOE is the mean spectral density $\rho_n(t)=\mathbb{E}\left\{\frac{1}{n}\sum_{i=1}^n \delta(t-\lambda_i)\right\}$ such that the mean number $\#(a,b)$ of GOE eigenvalues in any interval $[a,b]$ is given by $\#(a,b)=n\int_a^b \rho_n(t)\,dt$ . According to (\ref{GOE2})
this can be rewritten as
 \begin{equation}\label{GOE4}
\rho_{n,a}(t)= Z^{-1}_{n}(a)\int_{\mathbb{R}^N} \delta(t-\lambda_n)\,e^{-\frac{1}{2a}\sum_{i=1}^n \lambda_i^2} |\Delta_n(\Lambda)|\, d\Lambda\,,
\end{equation}
where we have used the symmetry of the integrand with respect to permutation of the variables $\lambda_i$.
Notice that the mean density for two different values of the variance parameter $a_1$ and $a_2$ satisfies a simple re-scaling property:
\begin{equation}\label{GOE4b}
\sqrt{\frac{a_1}{a_2}}\rho_{n,a_1}\left(t\sqrt{\frac{a_1}{a_2}}\right)=\rho_{n,a_2}\left(t\right), \quad \forall a_1>0,a_2>0\,.
\end{equation}
We can further use the decomposition of the Vandermonde factor:
\begin{equation}\label{GOE4c}
|\Delta_{n}(\Lambda)|=|\Delta_{n-1}(\Lambda)|\prod_{j=1}^{n-1}|\lambda_n-\lambda_j|
\end{equation}
  to bring (\ref{GOE4}) to the following form:
  \begin{equation}\label{GOE4a}
\rho_{n,a}(t)== Z^{-1}_{n}(a) e^{-\frac{1}{2a}t^2}\int_{\mathbb{R}^{N-1}}\, \,e^{-\frac{1}{2a}\sum_{i=1}^{n-1} \lambda_i^2} \, \prod_{j=1}^{n-1}|t-\lambda_j|\,|\Delta_{n-1}(\Lambda)|\, d\Lambda\,.
\end{equation}

A useful observation going back to \cite{Fyo04} is that the integral in the right-hand side of (\ref{GOE4a}) can be interpreted as being proportional to $\mathbb{E}\left\{|\det(t{\bf 1}-\tilde{H})|\right\}$, where the random $(n-1)\times (n-1)$ matrix $\tilde{H}$ is sampled with the same GOE measure (\ref{GOE1}) with the reduced size $n\to n-1$. This observation provides us with the following relation:
\begin{equation}\label{GOE5}
\int |\det(t{\bf 1}-\tilde{H})|\,d\mu^{(GOE)}_{n-1,a}(\tilde{H})=\frac{Z_{n}(a)}{Z_{n-1}(a)}e^{\frac{1}{2a}t^2}\rho_{n,a}(t).
\end{equation}
Importance of this relation becomes obvious after realizing that its left hand side is (up to the normalization factor $C_{n-1}(a)$) the same as
the matrix integral featuring in the right-hand side of (\ref{KRsph17}). Substituting (\ref{GOE5})
to (\ref{KRsph17}) one therefore gets after a straightforward rearrangement an expression for the mean number of
stationary points of any $N-$dimensional isotropic random field constrained to a sphere of radius $R$ in terms of the
mean eigenvalue density $\rho_{N,a}(t)$ of $N\times N$ GOE matrix with the variance parameter $a=2F''(R^2)$.
It is also convenient to change at this point $t\to t \sqrt{aN}$ and make use of the
re-scaling property of the eigenvalue density (\ref{GOE4b}) with the particular choice $a_1=a,\,a_2=1/N$: $\sqrt{aN}\rho_{N,a}(t\sqrt{aN})=\rho_{N,1/N}(t)\equiv
\rho_{N}(t)$.  The latter notation $\rho_{N}(t)$ stands for the mean eigenvalue density for the "standardized" GOE with the variance parameter $1/N$. In terms of such a density the mean number of stationary points on the sphere is given by
 \begin{equation}\label{GOE6}
\mathbb{E}\left\{{\cal N}_s\right\}=2N \left(\frac{R^2 F''(R^2)}{F'(R^2)}\right)^{N/2}
\end{equation}
\[
\times \sqrt{\frac{2F'(R^2)/R^2}{ F'(R^2)/R^2+F''(R^2)}}
 \int_{-\infty}^{\infty} e^{-N\frac{t^2}{2}\frac{F''(R^2)-F'(R^2)/R^2}{F''(R^2)+F'(R^2)/R^2}}\rho_{N}(t)\,dt.
\]
Finally, we notice that the above expression depends on a single parameter
\begin{equation}\label{GOE7a}
 B=\frac{F''(R^2)-F'(R^2)/R^2}{F''(R^2)+F'(R^2)/R^2}
\end{equation}
 which allows to
rewrite it in a rather compact form:
\begin{equation}\label{GOE7}
\mathbb{E}\left\{{\cal N}_s\right\}=4N \left(\frac{1+B}{1-B}\right)^{N/2}
\sqrt{1-B}\,\, G(B), \,\
\end{equation}
where we have defined
\begin{equation}\label{GOE7prim}
 G(B)= \int_{0}^{\infty} e^{-N\frac{t^2}{2}B}\rho_{N}(t)\,dt
\end{equation}
and  used that $\rho_{N}(t)=\rho_{N}(-t)$. As is well known, the mean density  $\rho_{N}(t)$ of the standardized GOE can be expressed for any $N$ in terms of the Hermite polynomials $H_j(x)$, see e.g. the Chapter 7.2 in \cite{Mehtabook}, in particular eq.(7.2.32). For example, for $N$ even we have
 \begin{equation}\label{GOEdens}
\rho_N(x)=\sqrt{N}\phi_N\left(\sqrt{N}x\right)-\sqrt{N+1)}\phi_{N-1}\left(\sqrt{N}x\right)\phi_{N+1}
\left(\sqrt{N}x\right)
\end{equation}
\[
+\frac{1}{2\sqrt{2}}\phi_{N-1}\left(\sqrt{N}x\right)\int_{-\infty}^{\infty} s(x-t)\phi_{N}\left(t\right)\,dt,
\]
 where $s(x)=x/|x|$ for $x\ne 0$ and $s(0)=0$ is the sign function and
 \begin{equation}
 \phi_j(x)=(2^jj!\sqrt{\pi})^{-1/2}e^{-\frac{x^2}{2}}H_j(x), \quad H_j(x)=e^{x^2}\left(-\frac{d}{dx}\right)^je^{-x^2}.
 \end{equation}
 We thus see that our formalism provides a closed form explicit expression for the mean number of stationary points for any finite positive integer $N$. In what follows we are going to apply it to understanding the random landscapes typical for the $p-$spin spherical model of spin glasses subject to a random magnetic field.

\subsection{Counting stationary points of the p-spin spherical spin glass model in a random magnetic field.}

The model was introduced and studied originally by Crisanti and Sommers \cite{CS} who generalized the simplest $p=2$ case investigated much earlier in \cite{KTJ}. In our notations we are going to be somewhat closer (but not always identical) to the paper \cite{Auf1} which addressed questions rather similar to the scope of the present lectures, yet from a somewhat different angle and with a different emphasis. The main object of study is the random energy function ("Hamiltonian") of $N$ interacting real degrees of freedom ("spins") denoted as $x_1,\ldots,x_N$ and taking arbitrary real values, provided  $\sum_{i=1}^Nx_i^2=N$. In other words, the vectors of allowed configurations $\x=(x_1,\ldots,x_N)$ are constrained to the surface of a sphere of the radius $R=\sqrt{N}$. Then to each allowed spin configuration one assigns the energy value:
\begin{equation}\label{pSG1}
V(x_1,\ldots,x_N)=-\sum_{1\le i_1, \ldots, i_p\le N}J_{i_1 \ldots i_p}x_{i_p} \cdots x_{i_1}-\sum_{i=1}^Nh_i x_i,   \end{equation}
where it is assumed that $p\ge 2$ is an integer, and so the model is characterized by $p-$spin interaction.
Here the summation goes over all the $p-$tuples $(i_1,i_2 \ldots, i_p)$ without restrictions
 and the interaction constants $J_{i_1 \ldots i_p}$ for {\it all} $p-$tuples
 are assumed to be independent mean zero real random Gaussian variables (so e.g. for $p=2$ the constants $J_{i_1i_2}$
  and $J_{i_2i_1}$ are independent for $i_1\ne i_2$) and the variances:
 \begin{equation}\label{pSG2}
\mathbb{E}\left\{\left(J_{i_1 \ldots i_p}\right)^2\right\}=\frac{J^2}{pN^{p-1}}
\end{equation}
We also treat the components $h_i, i=1,\ldots,N$ of the magnetic field ${\bf h}$ as random Gaussian variables, identically distributed with zero mean and the variance  $\langle h^2_i\rangle=\sigma^2$ and independent of all the coupling constants $J'$s. As a result, the energy $V(\x)$ is a random Gaussian function of spin variables $x_1,\ldots,x_N$ with zero mean and the covariance:
 \begin{equation}\label{pSG3}
\mathbb{E}\left\{V(\x)V(\x')\right\}=Nf\left(\frac{\x\cdot\x'}{N}\right), \quad f(u)=\frac{J^2}{p}u^p+\sigma^2 u
\end{equation}
We thus see that the random energy associated with the $p-$spin model is an example of the $N-$dimensional isotropic Gaussian field restricted to the sphere of the radius $R^2=N$. Hence all previously developed theory described by our main formulae (\ref{GOE7})-(\ref{GOE7a}) is fully applicable, with the main parameter $B$ specified to the present context given by
\begin{equation}\label{psG4}
B=\frac{J^2(p-2)-\sigma^2}{J^2p+\sigma^2}\in \left(-1, \frac{p-2}{p}\right].
\end{equation}

\subsubsection{Mean number of stationary points: large-$N$ asymptotics for a fixed value of $\sigma$.}
Actually, in the theory of spin glasses one is mainly interested in investigating the system's behaviour
in the so-called thermodynamic limit $N\to \infty$. We will now proceed to the corresponding asymptotic analysis of  (\ref{GOE7}). We will see that the asymptotic of the mean number of stationary points will be very different
for positive and negative  values of $B$, with the point $B=0$ corresponding to $\sigma=J\sqrt{p-2}$  playing a role of the critical point of a spin-glass phase transition.

To perform the asymptotic analysis for a given, $N-$independent value of $B$ we will use the well-known
asymptotic behaviour of the mean density $\rho_N(t)$ of the standardized GOE ensemble, see \cite{For1,Fyo04}:
\begin{equation}\label{psG5}
\rho_{N\gg 1}(t)=\left\{\begin{array}{cc} \frac{1}{\pi}\sqrt{2-t^2}, & 0<t<\sqrt{2}\\ \frac{1}{2\sqrt{\pi N}}
\frac{e^{-N\psi_{+}(t)}}{(t^2-2)^{1/4}\left(t+\sqrt{t^2-2}\right)^{1/2}}, & t>\sqrt{2}\end{array}\right.
\end{equation}
where
\begin{equation}\label{psG6}
\psi_{+}(t)=\frac{t}{2}\sqrt{t^2-2}-\ln{\frac{t+\sqrt{t^2-2}}{\sqrt{2}}}\,.
\end{equation}
Correspondingly, for the purpose of asymptotic analysis of (\ref{GOE7}) it is reasonable to decompose
$G(B)=G_<(B)+G_>(B)$ where we have defined
\begin{equation}\label{psG7}
G_<(B)= \int_{0}^{\sqrt{2}} e^{-N\frac{t^2}{2}B}\rho_{N}(t)\,dt, \quad G_>(B)= \int_{\sqrt{2}}^{\infty} e^{-N\frac{t^2}{2}B}\rho_{N}(t)\,dt.
\end{equation}
In the first integral we can replace $\rho_{N}(t)$ for $N\gg 1$ with its asymptotic semicircular form $\frac{1}{\pi}\sqrt{2-t^2}$ according to (\ref{psG5}),
which after replacing $t=\sqrt{2}\cos{\theta}$ allows, without further approximations, to express the integral in terms of the modified Bessel function $I_0(z)=\frac{1}{\pi}\int_0^{\pi}e^{z\cos{\phi}}d\phi$. We have
\begin{equation}\label{psG8}
G_<(B)= -e^{\gamma}\frac{d}{d\gamma}\left(e^{-\frac{\gamma}{2}} I_0\left(\frac{\gamma}{2}\right)\right), \quad \gamma=-N\,B\,.
\end{equation}
For any fixed $B<0$ and $N\gg 1$ we have $\gamma\gg 1$, hence using the asymptotic form $I_0\left(\frac{\gamma}{2}\right)|_{|\gamma|\gg 1}\approx \frac{1}{\sqrt{\pi |\gamma|}}e^{\frac{|\gamma|}{2}}$
yields in this case the exponentially growing behaviour:
\begin{equation}\label{psG9}
G_<(B<0)|_{N\gg 1}\approx \frac{e^{N\,|B|}}{\left(\pi N^3|B|^3\right)^{1/2}}.
\end{equation}
Similarly, for any fixed $B>0$ and $N\gg 1$ we have $\gamma\to -\infty$ which then results in a powerlaw
decaying asymptotic:
\begin{equation}\label{psG10}
G_<(B>0)|_{N\gg 1}\approx \frac{1}{\left(\pi N B\right)^{1/2}}.
\end{equation}
Now we similarly proceed to extracting the asymptotic behaviour of $G_>(B)$ by considering the integral
\begin{equation}\label{psG11}
G_>(B)|_{N\gg 1}\approx \frac{1}{2\left(\pi N\right)^{1/2}}\int_{\sqrt{2}}^{\infty}
\frac{e^{-NL_{>}(t)}}{(t^2-2)^{1/4}\left(t+\sqrt{t^2-2}\right)^{1/2}}\, dt,
\end{equation}
where
\begin{equation}\label{psG12}
L_{>}(t)=\frac{B}{2}t^2+\frac{t}{2}\sqrt{t^2-2}-\ln{\frac{t+\sqrt{t^2-2}}{\sqrt{2}}}.
\end{equation}
Note that the integral is convergent for any allowed $B$  as $L_{>}(t\gg 1)\approx \frac{(1+B)}{2}t^2$ and $B>-1$ according to (\ref{psG4}). After straightforward algebra we find $\frac{d}{dt}L_{>}(t)=Bt+\sqrt{t^2-2}$, hence only for $B<0$ there exists an  extremum of $L_{>}(t)$ at $t=t_*=\sqrt{\frac{2}{1-B^2}}>\sqrt{2}$. Further differentiation gives $\frac{d^2}{dt^2}L_{>}(t_*)=-\frac{1-B^2}{B}>0$ so this is indeed the minimum and the integral is dominated by the vicinity of $t=t_*$. We further have $L_{>}(t_*)=-\frac{1}{2}\ln{\frac{1-B}{1+B}}$ yielding
\begin{equation}\label{psG13}
G_>(B<0)|_{N\gg 1}\approx \frac{1}{2N}\frac{1}{\sqrt{1-B}} e^{N\frac{1}{2}\ln{\frac{1-B}{1+B}}}.
\end{equation}
On the other hand for $B>0$ we have $\frac{d}{dt}L_{>}(t)>0$ in the whole domain $t>\sqrt{2}$, and the integral
is dominated by the vicinity of the lower limit $t=\sqrt{2}$. The straightforward evaluation then gives the asymptotic:
\begin{equation}\label{psG13a}
G_>(B>0)_{N\gg 1}\approx \frac{1}{4\sqrt{\pi N}}\frac{\Gamma(3/4)}{(NB)^{3/4}} e^{-NB}.
\end{equation}
Now, by comparing (\ref{psG13}) with (\ref{psG9}) and taking into account that
$\frac{1}{2}\ln{\frac{1+|B|}{1-|B|}}>|B| , \forall |B|\ne 0$ we conclude $G_>(B<0)\gg G_<(B<0)$ as long as ${N\gg 1}$. Hence to the leading order $G(B<0)|_{N\gg 1}$ is equal to $G_>(B<0)|_{N\gg 1}$
and substituting (\ref{psG13}) for $G(B)$ in (\ref{GOE7}) we arrive finally at
\begin{equation}\label{psG14}
\lim_{N\to \infty}\mathbb{E}\left\{{\cal N}_s\right\}=2, \quad   \sigma>\sigma_c=J\sqrt{p-2}.
\end{equation}
This may look as a somewhat surprising result. Indeed, basic topological arguments predict that any smooth non-constant function must achieve both its minimum and its maximum on a sphere, hence ${\cal N}_s=2$ is the absolute minimum number of stationary points of any field in such a setting meaning there exists only one minimum and only one maximum in the corresponding landscape. We just see that any magnetic field exceeding the threshold  $\sigma_c$ by whatever small but $N-$independent amount results in total trivialization of the topology of the landscape in the
thermodynamic limit $N\to \infty$. In particular, for the simplest case $p=2$ {\it any} nonzero $N-$independent field leads to such an effect, the fact first noticed in \cite{CD1} and discussed in much detail in \cite{FLD2013}.

Let us consider now $0<\sigma<\sigma_c$ which corresponds to $B>0$.  By comparing (\ref{psG10}) with (\ref{psG13a}) we conclude that $G_>(B>0)\ll G_<(B>0)$ as long as $N\gg 1$, so that asymptotically $G(B>0)_{N\gg 1}$ should be replaced with (\ref{psG10}) which upon substituting to (\ref{GOE7}) leads to
\begin{equation}\label{psG15}
\mathbb{E}\left\{{\cal N}_s\right\}|_{N\gg 1}\approx 4N^{1/2}\sqrt{\frac{1+B}{\pi B}}\, e^{N\frac{1}{2}\ln{\frac{1+B}{1-B}}}, \quad   \sigma<\sigma_c=J\sqrt{p-2}
\end{equation}
As $\frac{1}{2}\ln{\frac{1+B}{1-B}}>0$ for $B>0$ we conclude that for any $\sigma<\sigma_c$ the number of stationary points in the random energy landscape is exponentially big in the thermodynamic limit, and the corresponding exponent known as the "cumulative complexity" vanishes linearly with $\sigma_c-\sigma$ when $\sigma\to \sigma_c$. Such a change in the landscape topology will have implications for the thermodynamic behaviour of the model, accompanied with such spin glass effects like strong ergodicity breaking. In the standard physical language this change is said to be reflected by the (one-step) replica symmetry breaking \cite{CS}.

\subsubsection{Mean number of stationary points: large-$N$ asymptotic close to the transition point $\sigma=\sigma_c$.}

After we have understood the global picture the next natural step is to try to investigate in more detail
the vicinity of the threshold value $\sigma=\sigma_c$ of the magnetic field variance, or equivalently the vicinity of $B=0$.
In this way we should obtain a detailed picture of the crossover between the two drastically different types of behaviour in the mean number of stationary points. As we have seen from the previous analysis, the change
between the two regimes is technically due to the change of the relevant behaviour of the mean eigenvalue density $\rho_N(t)$ described in the two lines of the formula (\ref{psG5}) for $t<\sqrt{2}$ and $t>\sqrt{2}$, respectively. As is well known, there is a smooth crossover between the two types of behaviour in the so-called "edge scaling" regime taking place in a small vicinity, of the widths $N^{-2/3}$, of the spectral edge $t=\sqrt{2}$. More precisely, introducing the scaling $t=\sqrt{2}\left(1+\frac{\zeta}{2N^{2/3}}\right)$ and considering $\zeta$ to be of the order of unity
 one finds that $\rho_N(t)\approx N^{-1/3}\sqrt{2}\,\rho_{edge}(\zeta)$ where explicit expression for $\rho_{edge}(\zeta)$ is given by\cite{For1993}
 \begin{equation}\label{edgedens}
  \rho_{edge}(\zeta)=\left[Ai'(\zeta)\right]^2-\zeta\left[Ai(\zeta)\right]^2+\frac{1}{2}Ai(\zeta)
 \left(1-\int_{\zeta}^{\infty}
 Ai(\eta)\,d\eta\right),
\end{equation}
where $Ai(\zeta)=\frac{1}{2\pi i}\int_{\Gamma}^{}e^{\frac{v^3}{3}-v\zeta}\, dv$ is the Airy function solving the differential equation $Ai''(\zeta)-\zeta Ai(\zeta)=0$.

\begin{figure}[htpb]
  \centering
  \includegraphics[width=0.8\textwidth]{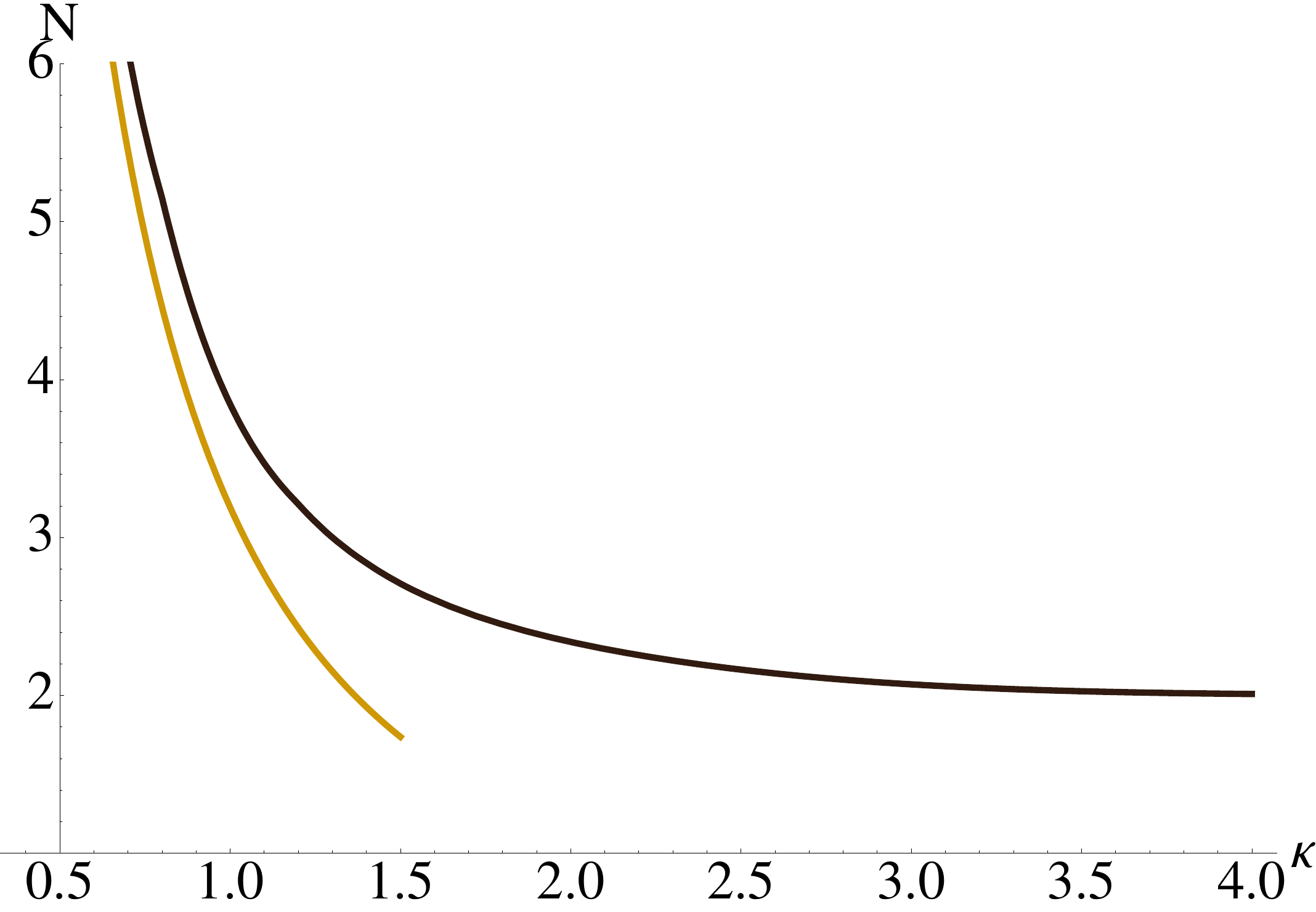}
  \caption{ Mean number N$\equiv \lim_{N\to \infty}\mathbb{E}\left\{{\cal N}_s\right\}$ of stationary points as a function of $\kappa=N^{1/3}|B|$ in the "edge" critical scaling regime $\sigma>\sigma_c$ such that $\sigma-\sigma_c\sim N^{-1/3}$, from the formula (\ref{psG16}). The asymptotic
  formula for small $\kappa$, Eq. (\ref{totalnumber4}) is also indicated as the lower curve. For $\kappa \to \infty$ the mean number converges to the minimal possible value $2$ (see the text). }
\label{fig:fig2}
\end{figure}

It turns out that such a spectral crossover induces the existence of the critical crossover in the counting function $\mathbb{E}\left\{{\cal N}_s\right\}$, taking place for  $\sigma>\sigma_c$ such that $\sigma-\sigma_c\sim N^{-1/3}$. Namely, introducing the scaled value $B=-\frac{\kappa}{2 N^{1/3}}$ and considering $\kappa>0$ to be of the order of unity it is a straightforward exercise to show that (\ref{GOE7}) implies that
\begin{equation}\label{psG16}
\lim_{N\to \infty}\mathbb{E}\left\{{\cal N}_s\right\}= 4e^{-\kappa^3/24}\int_{-\infty}^{\infty} e^{\frac{\kappa}{2}\zeta}\rho_{edge}(\zeta)\,d\zeta, \quad \kappa=2N^{1/3}|B|, B<0\,.
\end{equation}
 We see that $\lim_{N\to \infty}\mathbb{E}\left\{{\cal N}_s\right\}$ always remains of the order of unity. This function is plotted in Fig. \ref{fig:fig2} borrowed from
 \cite{FLD2013}.

  As a check of consistency let us first verify that (\ref{psG16}) for $\kappa\to \infty$  approaches the limit
   $\mathbb{E}\left\{{\cal N}_s\right\}=2$ as predicted by (\ref{psG14}).
  The behaviour of $\mathbb{E}\left\{{\cal N}_s\right\}$ for $\kappa\gg 1$  is controlled by $\zeta\to \infty$ asymptotic behaviour of the spectral density:
 \begin{equation}\label{edgedensa}
 \rho_{edge}(\zeta\to +\infty)\approx \frac{1}{2} Ai(\zeta) \approx
   \frac{1}{4\sqrt{\pi} \zeta^{1/4}}\exp{\left\{-\frac{2}{3}\zeta^{3/2}\right\}}\,.
\end{equation}
Using this asymptotic we then find
\begin{equation}\label{totalnumber5}
\lim_{N\to \infty}\mathbb{E}\left\{{\cal N}_s\right\}|_{\kappa\gg 1}\approx \frac{e^{-\kappa^3/24}}{\sqrt{\pi}}\int_{0}^{\infty} e^{-\frac{2}{3}\zeta^{\frac{3}{2}}+\frac{\kappa}{2}\zeta}\, \frac{1}{\zeta^{1/4}} d\zeta
\end{equation}
\[
=\frac{e^{-\kappa^3/24}\kappa^{3/2}} {\sqrt{\pi}}\int_{0}^{\infty} e^{-\kappa^3\left(\frac{2}{3}u^{\frac{3}{2}}-\frac{u}{2}\right)}\, \frac{du}{u^{1/4}},
\]
where we have made a substitution $\zeta=u\,\kappa^2$ to make it evident that the integral in the limit $\kappa\gg 1$ can be evaluated by the Laplace method around the stationary point $u=1/4$\footnote{Alternatively we can use (\ref{edgedensa}) and the identity $\int_{-\infty}^{+\infty} d\zeta Ai(\zeta) e^{\frac{\kappa}{2} \zeta} =e^{\frac{\kappa^3}{24}}$ for any $\kappa \geq 0$.}. A straightforward calculation
then yields the value $\lim_{\kappa\to \infty}\lim_{N\to \infty}\mathbb{E}\left\{{\cal N}_s\right\}=2$ as predicted by (\ref{psG14}). This is the minimal possible value implying the total trivialization of the landscape topology, and existence of a single  minimum-maximum pair in the landscape.

 On the other hand, the behaviour of $\lim_{N\to \infty}\mathbb{E}\left\{{\cal N}_s\right\}$ for small values of the scaling parameter $0<\kappa\ll 1$  is obviously controlled by $\zeta\to -\infty$ asymptotics of the edge eigenvalue density:
\begin{equation}\label{edgedensaleft}
 \rho_{edge}(\zeta\to -\infty)\approx \frac{\sqrt{|\zeta|}}{\pi} , \quad
\end{equation}
 which implies
\begin{equation}\label{totalnumber4}
 \mathbb{E}\left\{{\cal N}_s\right\}|_{0<\kappa\ll 1}\approx 4\int_{-\infty}^{0}e^{\frac{\kappa}{2}\zeta}\, \frac{\sqrt{|\zeta|}}{\pi} d\zeta = \frac{4\sqrt{2}}{\sqrt{\pi} \kappa^{3/2}} \gg 1.
\end{equation}

 This result clearly indicates a trend towards anticipated drastic growth of the mean number of stationary points for $\kappa\to 0$. However the formula (\ref{totalnumber4}) does not yet match $B\to 0$ behaviour of  $\mathbb{E}\left\{{\cal N}_s\right\}$ from the side of positive $B>0$, see (\ref{psG15}), hence indicating the existence of yet another crossover critical regime. The latter actually takes place at $B\sim N^{-1}\ll N^{-1/3}$ where $B$ can be of any sign. Indeed, rescaling now $B=-\gamma/N$ and allowing $N\to \infty$ keeping $\gamma$ finite one find that (\ref{GOE7}) is reduced to:
\begin{equation}\label{totalnumber2}
 \lim_{N\to \infty}\frac{\mathbb{E}\left\{{\cal N}_s\right\}}{2N}= 2e^{-\gamma}\int_{0}^{\sqrt{2}}\sqrt{2-t^2}\,
e^{\frac{\gamma}{2} t^2} \,\frac{dt}{\pi}=-2\frac{d}{d\gamma} \left(e^{-\frac{\gamma}{2}}I_0\left(\frac{\gamma}{2}\right)\right).
\end{equation}
 We see that in such a regime the total number of stationary points in our random landscape is of the order of $N$, and this provides a sought for link between the value of the order of unity typical
for the "edge scaling" regime for $B<0$, and  the exponential in $N$ behaviour in the regime $B>0$.
In particular, exploiting the asymptotic $I_0(|z|\gg 1)\sim e^{|z|}/\sqrt{2\pi |z|}$ it is easily checked that the above expression tends for $\gamma\gg 1$ to  $ \frac{1}{\sqrt{\pi}}\gamma^{-3/2}$ which precisely matches the formula (\ref{totalnumber4}). On the other hand, for
$\gamma\to -\infty$ it tends to $2e^{|\gamma|}/\sqrt{\pi |\gamma|}$ which exactly matches the $B\ll 1$ limit of
(\ref{psG15}). This means we have now the complete picture of the crossover between the two phases.

\section{The mean number of minima for isotropic Gaussian random fields on a sphere}

After a considerable effort invested in understanding the mean number of all stationary points for isotropic Gaussian random fields on a sphere it is relatively easy to perform similar analysis for the mean number of minima $\mathbb{E}\left\{{\cal N}_m\right\}$. Minima are usually considered to be especially important for statistical mechanics applications as the low-temperature behaviour should be controlled by the deepest minima in the landscape.  The starting expression for the mean number of minima (\ref{KRMin})
differs from its counterpart (\ref{KR}) for all stationary points by the presence of an indicator function
 ensuring the positivity of the Hessian. As the latter condition depends only on the eigenvalues of the Hessian
 it is immediately clear that for isotropic Gaussian fields on the $R-$sphere the formula (\ref{KRsph17}) should be replaced with
\begin{equation}\label{min1}
\mathbb{E}\left\{{\cal N}_m\right\}=2\frac{V_{N-1}(R)}{2^{(N-1)/2}\sqrt{b}}\,\frac{1}{[2\pi F'(R^2)]^{(N-1)/2}}\,\frac{1}{[2\pi F''(R^2)]^{N(N-1)/4}}
\end{equation}
\[
\times \int_{-\infty}^{\infty}e^{-\frac{t^2}{2b}} \frac{dt}{\sqrt{2\pi}}\,\int \, e^{-\frac{1}{2a}{\small \mbox{Tr}H^2}} \,\det{(t{\bf 1}-H)}\, \theta(t{\bf 1}-H)\, \, dH,
\]
where $H$ is $(N-1)\times (N-1)$ real symmetric matrix. Understanding this type of integral
 again requires the use of tools borrowed from the Random Matrix Theory (RMT), but this time is related not to
the mean eigenvalue density of GOE but to the probability density of its maximal eigenvalue. Namely, define for $n\times n$ GOE (\ref{GOE1}) the random variable  $\lambda_{max}=\max\{\lambda_1,\ldots,\lambda_n\}$. We start with the obvious identity
\begin{equation}\label{min2}
{\cal F}_{n,a}(t)=\mbox{Prob}\{\lambda_{max}\le t\}=\sum_{i=1}^n \mbox{Prob}\{(\lambda_{max}=\lambda_i)\bigcap (\lambda_i\le t)\},\,
\end{equation}
which by the fact that the joint probability density of all eigenvalues is permutation-invariant can be written as
\begin{equation}\label{min2a}
{\cal F}_{n,a}(t)=n\,\mbox{Prob}\{(\lambda_{max}=\lambda_n)\bigcap (\lambda_n\le t)\}.
\end{equation}
Using the indicator functions $\theta(x)$ it can be further written as the expectation
\begin{equation}\label{min3}
{\cal F}_{n,a}(t)=n\,\mathbb{E}\left\{\theta(t-\lambda_n)\prod_{j=1}^{n-1} \theta(\lambda_n-\lambda_j)\right\}
\end{equation}
or equivalently, exploiting (\ref{GOE2}) and (\ref{GOE4c}) as
\begin{equation}\label{min4}
{\cal F}_{n,a}(t)=n\,Z^{-1}_{n}(a)\int_{-\infty}^{t}d\lambda_n\int_{-\infty}^{\lambda_n}d\lambda_{n-1} \int_{-\infty}^{\lambda_n}d\lambda_{n-2}\cdots
\int_{-\infty}^{\lambda_n}d\lambda_1
\end{equation}
\[
\times e^{-\frac{1}{2a}\sum_{i=1}^{n-1} \lambda_i^2}
 |\Delta_{n-1}(\Lambda)|\, e^{-\frac{1}{2a} \lambda_n^2} \, \prod_{j=1}^{n-1} |\lambda_n-\lambda_j|.
\]
Differentiating the latter over $t$ yields the relation
\begin{equation}\label{min5}
\frac{d}{dt}{\cal F}_{n,a}(t)=n \, \frac{e^{-\frac{1}{2a} t^2}}{Z_{n}(a)} \int_{-\infty}^{t}d\lambda_{1} \cdots
\int_{-\infty}^{t}d\lambda_{n-1}\,\prod_{j=1}^{n-1} (t-\lambda_j) \,  e^{-\frac{1}{2a}\sum_{i=1}^{n-1} \lambda_i^2}
 |\Delta_{n-1}(\Lambda)|
\end{equation}
\begin{equation}\label{min6}
=n \, e^{-\frac{1}{2a} y^2}\frac{Z_{n-1}(a)}{Z_{n}(a)} \int \, \det{(t{\bf 1}-\tilde{H})}\, \theta(t{\bf 1}-\tilde{H})\, \, d\mu^{(GOE)}_{n-1,a}(\tilde{H}),
\end{equation}
thus providing a relation between the density  $\frac{d}{dt}{\cal F}_{N,a}(t)$ of the largest GOE eigenvalue and the matrix integral entering the relation (\ref{min1}). Noting that the density  $\frac{d}{dt}{\cal F}_{N,a}(t)$ satisfies the same re-scaling identity (\ref{GOE4b}), and denoting the probability for the largest eigenvalue of the standardized GOE with the variance parameter $a=1/N$ simply as ${\cal F}_{N}(t)$ we arrive at the formula for the mean number of minima in the form fully analogous to (\ref{GOE7}):
\begin{equation}\label{min7}
\mathbb{E}\left\{{\cal N}_m\right\}=2 \left(\frac{1+B}{1-B}\right)^{N/2}
\sqrt{1-B}\,\, {\cal G}(B),
\end{equation}
where
\begin{equation}\label{min8}
 {\cal G}(B)= \int_{-\infty}^{\infty} e^{-N\frac{t^2}{2}B}\frac{d}{dt}{\cal F}_{N}(t)\,dt
\end{equation}
and the parameter $B$ was defined earlier in (\ref{GOE7a})\footnote{In fact it is easy to show following the same steps that replacing the probability density $\frac{d}{dt}{\cal F}_{N}(t)$ of the largest GOE eigenvalue $\lambda_{max}$ with the probability density of the $k-$th largest GOE eigenvalue yields the mean number of stationary points with the index (i.e. the number of negative eigenvalues of the Hessian) equal to $k-1$, see \cite{Auf1,Auf2}.}.

Again, to perform the asymptotic analysis for a given, $N-$independent value of $B$ we will use the
asymptotic behaviour of the  probability density $\frac{d}{dt}{\cal F}_{N}(t)$ for the standardized GOE ensemble investigated for various regimes in \cite{For1,BDG,DM,MV,BorotNadal11,BorotNadal12}, see \cite{MajSch} for an informal and informative introduction:
\begin{equation}\label{min9}
\frac{d}{dt}{\cal F}_{N}(t)_{N\gg 1}=\left\{\begin{array}{cc} \sim e^{-N^2\psi_{-}(t)+N\Phi_1(t)}, & t<\sqrt{2}\\ \frac{N^{1/2}}{2\sqrt{\pi}}
\frac{e^{-N\psi_{+}(t)}}{(t^2-2)^{1/4}\left(t+\sqrt{t^2-2}\right)^{1/2}}, & t>\sqrt{2}\end{array}\right.
\end{equation}
where
\begin{equation}\label{min10}
\psi_{-}(t)=\frac{t^2}{3}-\frac{t^4}{108}-\left(\frac{t^3}{108}+ \frac{5 t}{36}\right)\sqrt{t^2+6}-
\frac{1}{2}\ln\left[\frac{t+\sqrt{t^2+6}}{3 \sqrt{2}}\right]\,,
\end{equation}
and
\begin{equation}\label{min11}
\psi_{+}(t)=\frac{t}{2}\sqrt{t^2-2}-\ln{\frac{t+\sqrt{t^2-2}}{\sqrt{2}}}
\end{equation}
and the explicit expression for $\Phi_1(t)$ is rather long and can be found in \cite{BorotNadal11}, but turns out to be actually immaterial for our purposes apart from the fact that $\Phi_1(t)\propto (\sqrt{2}-t)^{3/2}$ when $t\to \sqrt{2}$. Correspondingly, for the purpose of asymptotic analysis of (\ref{min8}) it is reasonable to decompose
${\cal G}(B)={\cal G}_<(B)+{\cal G}_>(B)$ where we have defined
\begin{equation}\label{min12}
{\cal G}_<(B)= \int_{-\infty}^{\sqrt{2}} e^{-N\frac{t^2}{2}B}\frac{d}{dt}{\cal F}_{N}(t)\,dt, \quad {\cal G}_>(B)= \int_{\sqrt{2}}^{\infty} e^{-N\frac{t^2}{2}B}\frac{d}{dt}{\cal F}_{N}(t)\,dt.
\end{equation}
Noticing that asymptotically ${\cal G}_>(B)\approx N G_>(B)$, the results for ${\cal G}_>(B)$ can be simply read off from (\ref{psG13}) for $B<0$ and from (\ref{psG13a}) for $B>0$. At the same time ${\cal G}_<(B)$ is rather different from $G_<(B)$ so it requires a separate analysis.
It is actually evident that for $N\gg 1$ the integral for ${\cal G}_<(B)$ will be always dominated by the vicinity of the minimum of $\psi_{-}(t)$, which can be shown to happen precisely at the boundary of the integration domain $t=\sqrt{2}$.  Close to that value of $t$ the large-$N$ asymptotic behaviour of the probability density $\frac{d}{dt}{\cal F}_{N}(t)$ is actually known explicitly with a greater precision, including the correct pre-exponential factors\cite{BorotNadal11}:
\begin{equation}\label{exactasympt}
\frac{d}{dt}{\cal F}_{N}(t)_{N\gg 1}\approx A\, N^{\frac{47}{24}}\,\epsilon^{\frac{31}{16}}\,
\exp{\left[-N^2\frac{1}{6\sqrt{2}}\epsilon^3-N\frac{2^{1/4}}{3}\epsilon^{3/2}\right]}, \quad \epsilon=\sqrt{2}-t\ll 1
\end{equation}
where $A$ is a numerical constant given by $\ln{A}=-\frac{169}{96}\ln{2}+\frac{1}{2}\zeta'(1)$, with the derivative of the Riemann zeta-function given approximately by  $\zeta'(1)=-0.1654211437$. Substituting
(\ref{exactasympt}) to the expression (\ref{min12}) for ${\cal G}_<(B)$, it is easily checked that the integral is dominated by the domain where $\epsilon=\sqrt{2}-t\sim N^{-\frac{1}{2}}\ll 1$ so that the term of the order $\sim \epsilon^{3/2}$ in the exponential of (\ref{exactasympt}) does not eventually play any role. We correspondingly rescale $\epsilon=\frac{\tau}{\sqrt{2N}}$ and evaluate the integral over the variable $\tau$ by the saddle-point method. It turns out that for $B>0$ the contribution of ${\cal G}_<(B)$
to ${\cal G}(B)$ is dominant, whereas for $B<0$ it is subdominant. This finally yields the following asymptotic behaviour:
\begin{equation}\label{min13a}
{\cal G}(B)_{N\gg 1}\approx  A\, 2^{\frac{35}{16}}\, N^{-\frac{17}{36}}\, B^{\frac{23}{32}}\,e^{-NB+\frac{4\sqrt{2}}{3} N^{1/2}B^{3/2}} , \quad B>0
\end{equation}
whereas our previous considerations imply
\begin{equation}\label{min13b}
{\cal G}(B)_{N\gg 1}\approx \frac{1}{2}\frac{1}{\sqrt{1-B}} e^{N\frac{1}{2}\ln{\frac{1-B}{1+B}}} , \quad B<0
\end{equation}
 The last expression immediately yields the mean total number of the minima asymptotically
 equal exactly to unity for $\sigma>\sigma_c$.  At the same (\ref{min13a}) implies that for $\sigma<\sigma_c$ the number of minima behaves asymptotically as
\begin{equation}\label{min14a}
\mathbb{E}\left\{{\cal N}_m\right\}|_{N\gg 1} = C_N(B)\, e^{N\Sigma(B)}, \quad \Sigma(B>0)=\frac{1}{2}\ln{\frac{1+B}{1-B}}-B\,.
\end{equation}
where the factor is given explicitly by
\begin{equation}\label{min14b}
 C_N(B)=8 A\, 2^{\frac{3}{16}}\, N^{-\frac{17}{36}}\, B^{\frac{23}{32}}\,\sqrt{1-B}\, e^{\frac{4\sqrt{2}}{3} N^{1/2}B^{3/2}}
\end{equation}
Note that for any $B>0$ the "complexity of minima"  $\Sigma(B)$ is positive,
and vanishes as $\Sigma(B)\sim \frac{1}{3}B^3 \propto (\sigma_c-\sigma)^3$ when approaching the magnetic field threshold $\sigma_c$. This is a manifestation of the so-called third-order transition \cite{MajSch}.

The remaining task is to establish the behaviour of the mean number of minima in the scaling crossover region
 in the vicinity of the magnetic field threshold $\sigma=\sigma_c$. As is well-known, the distribution ${\cal F}_{N}(t)$ of the largest eigenvalue
 of the standardized GOE in the "edge scaling" regime $t=\sqrt{2}\left(1+\frac{\zeta}{2N^{2/3}}\right)$ tends to the famous Tracy-Widom distribution \cite{TW,TWrev}:
\begin{equation}\label{TWdistr1}
\lim_{N\to \infty}{\cal F}_{N}(t)|_{t=\sqrt{2}\left(1+\frac{\zeta}{2N^{2/3}}\right)}=F_{1}(\zeta)
\end{equation}
which has the known explicit representation
 \begin{equation}\label{TWdistr2}
F_{1}(\zeta)=\exp\left\{-\frac{1}{2}\int_{\zeta}^{\infty}q(x)\,dx-\frac{1}{2}\int_{\zeta}^{\infty}(x-\zeta)q^2(x)\,dx\right\}
\end{equation}
 in terms of the solution $q(x)$ of the Painleve II equation $\frac{d^2q}{dx^2}=xq+2q^3$ with the boundary condition $q(x\to \infty)\approx Ai(x)$.

Correspondingly, the density $\frac{d}{dt}F_N(t)$ tends to $\sqrt{2}N^{2/3}\frac{dF_1}{d\zeta}$.
 After introducing the familiar scaling $B=-\frac{\kappa}{2 N^{1/3}}$ and performing the limit  $N\to \infty$ in  (\ref{min7}) while assuming the parameter $\kappa$ to be of the order of unity we arrive at the following expression for the mean number of minima in the crossover regime:
\begin{equation}\label{totalnumberminima}
  \lim_{N\to \infty}\mathbb{E}\{{\cal N}_m\}=2e^{-\kappa^3/24}
 \int_{-\infty}^{\infty} e^{\frac{\kappa}{2} \zeta}F_1'(\zeta)  d\zeta,
\end{equation}
where $F_1'(\zeta)=\frac{dF_1}{d\zeta}$. Note that in contrast to (\ref{psG16}) the above expression is well-defined also for  $\kappa<0$ due to an appropriate decay of the density $F_1'(\zeta)$, see below.  For a positive large $\kappa\gg 1$ the integral is controlled by the right
tail of the Tracy-Widom distribution, which takes the form \cite{TWrev}:
\begin{equation}\label{tailTWright}
\frac{dF_1}{d\zeta}|_{\zeta\gg 1}\approx \rho_{edge}(\zeta \gg 1) \approx \frac{1}{2} Ai(\zeta \gg 1) \approx
\frac{1}{4\sqrt{\pi} \zeta^{1/4}}\exp{\left\{-\frac{2}{3}\zeta^{3/2}\right\}}
\end{equation}
Exploiting (\ref{tailTWright}) and (\ref{totalnumber5}) we see that $ \lim_{\kappa\to\infty}\lim_{N\to \infty}\mathbb{E}\{{\cal N}_m\}=1$ which perfectly matches the single minimum regime $\sigma>\sigma_c.$ It is curious to observe that
the mean number of minima at vanishing $\kappa=0$ can be exactly calculated:  $\lim_{N\to \infty}\mathbb{E}\{{\cal N}_m\}|_{\kappa= 0}=2[F_1(\infty)-F_1(-\infty)]=2$, that is precisely at the critical point we have
on average two different minima.

Finally, the behaviour at $\kappa \to -\infty$ is controlled by the left tail of the TW distribution \cite{TWrev}:
\begin{equation}\label{tailTWleft}
F_1{\zeta}|_{\zeta\to -\infty}=2^{\frac{49}{32}}\,A \frac{1}{|\zeta|^{1/16}}\exp{\left\{-\frac{1}{24}|\zeta|^{3}-\frac{1}{3\sqrt{2}}|\zeta|^{3/2}\right\}}.
\end{equation}
whose derivative actually coincides with the edge scaling limit of (\ref{exactasympt}). We conclude that the asymptotic behaviour of $\mathbb{E}\{{\cal N}_m\}$ in this situation can be estimated from the integral
\begin{equation}\label{totalnumberextrema}
  \lim_{N\to \infty}\mathbb{E}\{{\cal N}_m\}|_{\kappa\to -\infty}\propto e^{|\kappa|^3/24}
 \int_0^{\infty} \zeta^{2-\frac{1}{16}}\, e^{|\kappa|\zeta-\frac{1}{24}\zeta^{3}-\frac{1}{3\sqrt{2}}\zeta^{3/2}}  d|\zeta|.
\end{equation}
 Changing $|\zeta|\to |\kappa|^{1/2} |\zeta|$ and applying the standard Laplace method for $|\kappa|\gg 1$ yields for the integral a factor of the order of $\exp{ \left(\frac{4\sqrt{2}}{3} |\kappa|^{3/2}\right)}$ which is subleading in comparison with the factor $\exp{\left(|\kappa|^3/24\right)}$. Remembering $|\kappa|=2B N^{1/3}$ we see that the latter factor precisely matches the behaviour $\exp{\frac{1}{3}B^3}$ of (\ref{min14a}) at $B\to 0$, with the subleading factor matching the exponential factor in $C(N,B)$ in (\ref{min14b}). We thus conclude that the formula (\ref{totalnumberminima}) valid everywhere in the "edge" scaling region $|\sigma-\sigma_c|\sim N^{-1/3}$ indeed ensures a smooth matching between the value (\ref{min14a}) and the value unity, and in this way decribes a gradual reduction of the mean number of minima in the crossover from the regime with exponentially many minima to one with just a single minimum, completing the picture of the transition.

 \section{Topology trivialization transition in a stationary Gaussian landscape with confinement.}
In order to see which features revealed in our analysis of the stationary points and minima  of the $p-$spin spherical model have chance to be universal it is reasonable to compare our findings with similar features of another simple, yet nontrivial model of a random Gaussian landscape characterized by the energy function
\begin{equation}\label{fundef}
{\cal H}=\frac{\mu}{2}\sum_{k=1}^{N-1} x_k^2+V(x_1,...,x_{N-1}),
\end{equation}
where the curvature $\mu>0$ of the non-random confining parabolic potential is used to control the number of
 stationary points and $V({\bf x})$ is a random mean-zero Gaussian-distributed field characterized by a particular translational invariant covariance structure:
\begin{equation}\label{translandisotr}
\mathbb{E} \left\{V(\mathbf{x}) V(\mathbf{y})\right\} = N\: f\left(\frac{1}{2 N} (\mathbf{x}-\mathbf{y})^2\right),
\end{equation}
with $f(x)$ being any smooth function suitably decaying at infinity. The mean number of stationary points for such a landscape were investigated in \cite{Fyo04,FyoWi07,my2005}, and the mean number of minima in \cite{FyoWi07,FyNa12},
see also \cite{BD07}. We will outline the main features below, and add a detailed analysis of the topology trivialization picture for that case. Let us also add, that although fields characterized by the covariance (\ref{translandisotr}) are both {\it stationary} and {\it isotropic}, it can be shown (see the Appendix) that one may relax the condition of isotropy and consider a particular class of anisotropic stationary fields, with the results simply related to those valid for the isotropic case.

Following essentially the same route as for the spherical model one can relate the  mean number of stationary points
for the parabolically confined random landscape (\ref{fundef}) to the mean eigenvalue density $\rho_{N}(t)$ of the standardized GOE:
\begin{equation}\label{parab1}
\mathbb{E}\left\{{\cal N}_s\right\}^{(par)}= \frac{2^{(N+1)/2}}{N^{(N-3)/2}}\frac{e^{-\frac{N}{2}m^2}}{m^{N-1}}\,\Gamma\left(\frac{N}{2}\right)
\int_{-\infty}^{\infty}\rho_{N}(t)e^{-\frac{N}{2}\left(t^2-2\sqrt{2}mt\right)},
\end{equation}
where $m=\frac{\mu}{f''(0)}>0$ is the main dimensionless control parameter of the theory.
In particular, in the limit of large $N\gg 1$ the $N-$dependence of $\mathbb{E}\left\{{\cal N}_s\right\}$
is very different for $0<m<1$ and for $m>1$. Namely, using the asymptotic of the mean density of eigenvalues (\ref{psG5}) one can show that in the first case the integral is dominated by the domain of $0<t<\sqrt{2}$ in the support of the semicircular law, with the result:
\begin{equation}\label{parab2}
\mathbb{E}\left\{{\cal N}_s\right\}^{(par)}\approx  4\sqrt{N}{\pi} m\sqrt{1-m^2}\,e^{N\Sigma^{(par)}_s(m)},
\end{equation}
where the complexity of minima $\Sigma^{(par)}_s(m)=\frac{1}{2}(m^2-1)-\ln{m}$ is positive for $0<m<1$,
and vanishes for $m\to 1$. At the same time in the case $m>1$ one can show that the integral is dominated
by a saddle-point located at $t>\sqrt{2}$, with the result
\begin{equation}\label{parab3}
\lim_{N\to\infty}\mathbb{E}\left\{{\cal N}_s\right\}^{(par)}=1, \quad m>1\,,
\end{equation}
which is the absolute minimum of a possible number of stationary points in such a landscape due to topological reasons. So we see that qualitatively the picture is very close to one found in the $p-$spin spherical spin glass, with the
value $m=m_c=1$ playing the role of the phase transition threshold. This fact is also confirmed by statistical mechanics calculations within the replica formalism, see \cite{FySo2007}.

It is therefore natural to investigate in more detail the scaling vicinity of the transition threshold, where we expect the phenomenon of gradual topology trivialization
to take place. We also may anticipate that the vicinity of the point $t=\sqrt{2}$ in the integral (\ref{parab1}) should play the dominant role. To that end we find it convenient to introduce a parameter ${\cal B}=m-1$ vanishing at
the transition point, and also change the integration variable $t\to \sqrt{2}(1-v)$. We will also replace the $\Gamma-$factor in (\ref{parab1}) with its asymptotic approximation: $\Gamma(N/2)\approx 2\sqrt{\pi/N}e^{-N/2}(N/2)^{N/2}$. In this way we rewrite (\ref{parab1}) as
\begin{equation}\label{parab4}
\mathbb{E}\left\{{\cal N}_s\right\}^{(par)}\approx  2\sqrt{2}N(1+{\cal B})e^{N\left[{\cal B}-\frac{1}{2}{\cal B}^2-\ln{\left(1+{\cal B}\right)}\right]}
\int_{-\infty}^{\infty}\rho_{N}\left(\sqrt{2}(1-v)\right)e^{-N\left(v^2+2{\cal B}v\right)}.
\end{equation}
It is rather easy to see that if we consider ${\cal B}>0$ and scale ${\cal B}=\frac{\kappa}{2N^{1/3}}$
the integral for $N\to \infty$ will be dominated by the values $v=O(N^{-2/3})$ and we immediately get
\begin{equation}\label{parab5}
\lim_{N\to \infty}\mathbb{E}\left\{{\cal N}_s\right\}= 2e^{-\kappa^3/24}\int_{-\infty}^{\infty} e^{\frac{\kappa}{2}\zeta}\rho_{edge}(\zeta)\,d\zeta, \quad \kappa=2N^{1/3}{\cal B}, \,\, {\cal B}>0\,,
\end{equation}
where $\rho_{edge}(\zeta)$ was defined in (\ref{edgedens}). We see that we have arrived at the number of order of unity
which is exactly half of the corresponding result for the $p-$spin spherical model (\ref{psG16}). So in such an "edge scaling" regime the mean number of the stationary points for two models becomes identical up to a trivial factor.
In particular, for $\kappa\to 0$ we have again the divergence predicted by (\ref{totalnumber4}) which points to the existence of yet another scaling regime which interpolates smoothly between the $\kappa\to 0$ behaviour of (\ref{parab5}) and $1-m\ll 1$ limit of (\ref{parab2}) which when written in terms of ${\cal B}=-|{\cal B}|\gg 1$ reads:
\begin{equation}\label{parab5a}
\mathbb{E}\left\{{\cal N}_s\right\}^{(par)}\approx  4\sqrt{2N|{\cal B}|}e^{N{\cal B}^2}, \quad {\cal B}<0
\end{equation}
The above formula indicates that such a second regime takes place for $|{\cal B}|\sim N^{-1/2}$.  To see this we  change the integration variable $v\to |{\cal B}| q$ in (\ref{parab4}), subdivide the integral over $q$ into the sum over $(-\infty,0]$ and $[0,\infty$, set
 $|{\cal B}|=\gamma N^{-1/2}$ and consider $\gamma$ to be of order of unity when $N\gg 1$. In this way we get:
 \begin{equation}\label{parab6}
\mathbb{E}\left\{{\cal N}_s\right\}^{(par)}\approx  2\sqrt{2} N^{1/2}\,|\gamma|\,\left(I_{-}(\gamma)+I_{+}(\gamma)\right),
\end{equation}
where
\begin{equation}\label{parab7}
I_{\pm}(\gamma)=\int_0^{\infty} \rho_{N}\left(\sqrt{2}(1\mp \frac{|\gamma|}{\sqrt{N}}q)\right)\,
e^{-\gamma^2\left(q^2\pm 2q s(\gamma)\right)}\,dq
\end{equation}
with $s(x)=x/|x|$ for $x\ne 0$ and $s(0)=0$ is the sign function.
Using asymptotic of the mean density of eigenvalues (\ref{psG5}) one can show that $I_{-}(\gamma)\ll I_{+}(\gamma)$ for any sign of $\gamma$.
Further using the semicircular law to approximate the mean density as $\rho_{N\gg 1}\left(\sqrt{2}\left(1-\frac{|\gamma|q}{N^{1/2}}\right)\right)\approx \frac{2\sqrt{q|\gamma|}}{\pi N^{1/4}}$ for $q>0$ of the order of unity, we arrive to the following expression:
 \begin{equation}\label{parab8}
\lim_{N\to \infty} N^{-1/4}\mathbb{E}\left\{{\cal N}_s\right\}^{(par)}|_{{\cal B}=\gamma N^{-1/2}} = \frac{4\sqrt{2}}{\pi} \,|\gamma|^{3/2}\,\int_0^{\infty}  \sqrt{q}
e^{-\gamma^2\left(q^2+2q s(\gamma)\right)}\,dq.
\end{equation}
This is the precise crossover formula analogous to (\ref{totalnumber2}) for the $p-$spin spherical model.
Indeed, on one hand it is easy to check that the limit $\gamma\to+\infty$ yields
the value $N^{1/4}/\left(\sqrt{\pi}\gamma^{3/2}\right)$ which precisely matches the $\kappa\to 0$ tail (\ref{totalnumber4}) in (\ref{parab5}).
On the other hand, in the limit $\gamma\to-\infty$ we reproduce the formula (\ref{parab5a}), thus perfectly describing the whole critical crossover. We see that in the parabolic confined potential the total number of stationary points in the crossover is always of the order of $N^{1/4}$ rather than $N$ as it was in the $p-$spin spherical model,
but otherwise there is obvious qualitative similarity between the two transitions.

Turning finally our attention to the mean number of minima for the parabolically confined landscape, we now understand that the general formula can be obtained from that for the total number of stationary points by replacing the mean density $\rho_N(t)$ with the the probability density for the largest eigenvalue of the standardized GOE with the variance parameter $a=1/N$ denoted $\frac{d}{dt}{\cal F}_{N}(t)$. The straightforward asymptotic analysis then yields
that $\lim_{N\to \infty}\mathbb{E}\left\{{\cal N}_m\right\}^{(par)}=1$ for $m>m_c=1$, whereas for $m<m_c$ it is given asymptotically by
\begin{equation}\label{min14par}
\mathbb{E}\left\{{\cal N}_m\right\}^{(par)}|_{N\gg 1} \sim e^{N\Sigma({\cal B})}, \quad \Sigma(m<1)={\cal B}-\frac{1}{2}{\cal B}^2-\ln{(1+{\cal B})}, \, \, {\cal B}=m-1\,.
\end{equation}
As expected, for any ${\cal B}>0$ the number of minima is exponentially big with positive complexity  $\Sigma({\cal B})>0$ which vanishes as $\Sigma({\cal B})\sim \frac{1}{3}B^3 \propto (m_c-m)^3$ when approaching the thereshold value of the confinement curvature. This is again manifestation of the third-order nature of the glass phase transition \cite{MajSch}. In fact, for the present model one is able to show that the free energy difference between the two phases has the same cubic singularity at $m\to m_c$ \cite{FySo2007}, so the meaning of the third order phase transition is fully justified also thermodynamically.

Moreover, after introducing the familiar scaling ${\cal B}=-\frac{\delta}{ N^{1/3}}$ and performing the limit  $N\to \infty$ in  (\ref{min7}) while assuming the parameter $\delta$ to be of the order of unity we arrive at the expression for the mean number of minima for the model with the parabolic confinement in the edge scaling regime in terms of the Tracy-Widom density $F_1'(\zeta)$
\begin{equation}\label{totalnumberminimapar}
  \lim_{N\to \infty}\mathbb{E}\{{\cal N}_m\}^{(par)}=2e^{-\delta^3/3}
 \int_{-\infty}^{\infty} e^{\delta \zeta}F_1'(\zeta)  d\zeta
\end{equation}
 which after the identification $\delta=\kappa/2$ is {\it identical} to the corresponding expression (\ref{totalnumberminima}) for the $p-$spin spherical model.

  In fact, that formula was first derived in \cite{FyNa12}, though was presented there from a somewhat different angle. Namely, having in mind to describe the smooth crossover between
 the two regimes for large but finite $N\gg 1$ it was suggested to replace the exact integral (\ref{totalnumberminimapar}) with its approximate value calculated by the Laplace method:
\begin{equation}\label{totalnumberminimaparapprox}
  \mathbb{E}\{{\cal N}_m\}^{(par)}\approx {\cal N}(\delta) = 2e^{-\delta^3/3}  e^{\delta\zeta_*}F_1'(\zeta_*) \sqrt{ \frac{2\pi}{-\frac{d^2}{d\zeta^2}\left[\ln{F_1'(\zeta_*)}\right]_{\zeta_*}}},
\end{equation}
where $\zeta_*$ is found from solving the saddle-point equation: $-\frac{d}{d\zeta}\left[\ln{F_1'(\zeta)}\right]_{\zeta_*}=\delta$. As the figure taken from \cite{FyNa12} shows
the subcritical $m>1$ and the supercritical $m<1$ regimes are indeed matched very smoothly by the above expression:

\begin{figure}[ht!]
\includegraphics[width=0.8\textwidth]{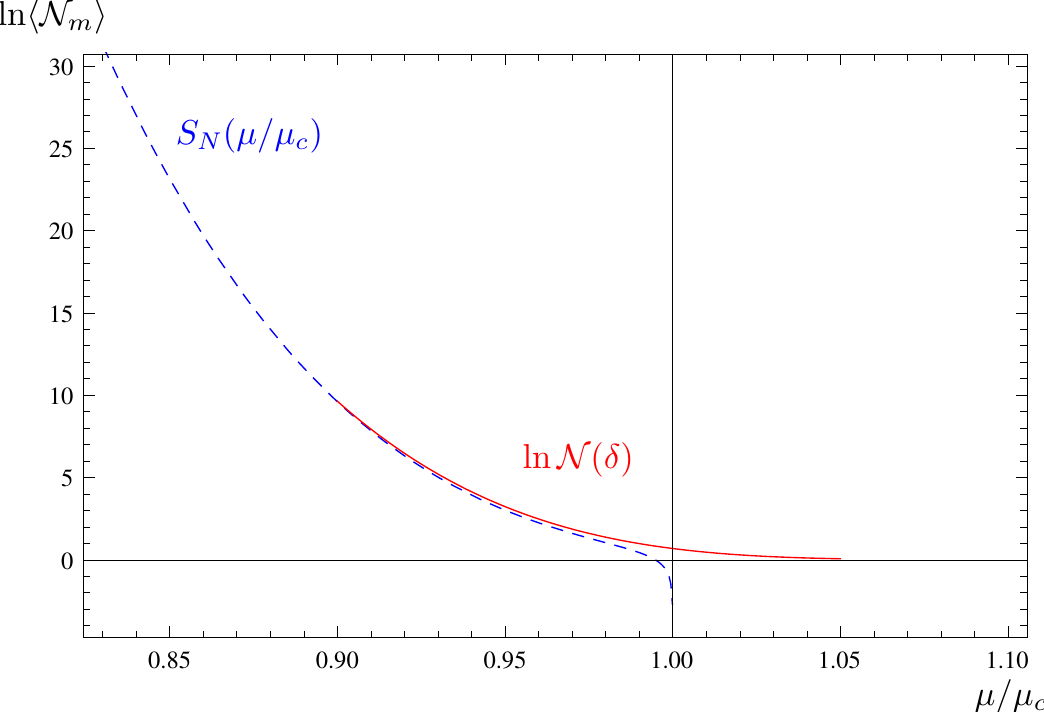}
\caption{Plot of the shape of the (log of the) mean number of minima $\langle \mathcal{N}_m\rangle\equiv \mathbb{E}\{{\cal N}_m\}^{(par)}$ for $N=10000$ as a function of $m=\mu/\mu_c$
(a) for $\mu < \mu_c$:  $\langle \mathcal{N}_m\rangle \sim e^{S_N}, S_N=N\Sigma$ from (\ref{min14par})(blue dashed line)
(b) in the transition region $\frac{\mu}{\mu_c}=1+\delta N^{-1/3}$ the graph is given by ${\cal N}(\delta)$ from (\ref{totalnumberminimaparapprox})  (red solid line)
(c) for $\mu >\mu_c$: $\langle \mathcal{N}_m\rangle \sim 1$ (black solid line)}
\label{fig}
\end{figure}

\section{Discussion and Conclusions}

We have presented a rather detailed theory of counting the mean number of stationary points and minima for high-dimensional random fields which are statistically isotropic or translationally invariant. Then we have applied it to understanding the process of gradual topology
trivialization for random energy landscapes for two particular models in the vicinity of the threshold of a zero-temperature glass-like transition, which at the level of statistical mechanics is known to be accompanied by the one-step spontaneous replica symmetry breaking.
On these two examples we revealed universality of the terminal stage of topology trivialization scenario which occurs in the vicinity of the threshold, with its widths scaling as $N^{-1/3}$. That stage, which is the only existing crossover regime for the number of landscape minima, is intimately related to the "edge scaling" behaviour of the GOE eigenvalues, in particular to the famous Tracy-Widom distribution. If however one is interested in totality of all the stationary points, the number displays yet another crossover in the region whose widths scales as $N^{-1}$ for the $p-$spin model and $N^{-1/2}$ for the parabolically confined case. Here the universality is less pronounced, and the crossover functions are different for the two models, though many qualitative features of the crossover look similar.

The following remark is appropriate here. As is easy to see for the simplest case $p=2$ spin glass the critical value of the magnetic field is zero: $\sigma_c=0$. This can be interpreted as the fact that $p=2$ case without any magnetic field is actually a kind of "critical" from the point of view of the spin glass transition. It is well-known that although thermodynamics of the $p=2$ model is simple and does not show such prominent features as replica-symmetry breaking or strong ergodicity breaking\cite{KTJ,CD}, dynamics is rich and has clear features of aging \cite{CD,ZKH,BDG}. That richness is attributed
to a relatively complicated energy landscape topology due to presence of $2N$ stationary points in the landscape,
see \cite{FLD2013} for a more detailed discussion. It is therefore tempting to conjecture that for a general $p>2$ similar features of aging should be seen everywhere in the second "critical scaling" regime $\sigma-\sigma_c\sim 1/N$ where the topology of the energy landscape seems to be very similar to $p=2$ case with zero (or small, $\sigma\sim N^{-1/6}$  \cite{FLD2013}) random magnetic fields. To extend this picture to threshold in the number of minima with changing the energy level of the random field as well as to the scaling vicinity of the finite-temperature thermodynamic transition  should be possible using the so-called TAP approach (see \cite{Auf1} and references therein for $p-$spin and \cite{spinglass2,spinglass2a} for the Sherrington-Kirkpatrick models) and deserves further investigations.
Note to this end the role of Tracy-Widom distribution in
the fluctuations of the temperature in the Sherrington-Kirkpatrick model \cite{Castelliana}.
On the other hand, the question of characterizing the fluctuation of number of critical points for high-dimensional landscapes remains largely open (see however
some steps towards addressing this problem in \cite{{KleinAgam}}). Here the application of recently proposed numerical algorithms which are able to find all the stationary points of high-dimensional random landscapes of polynomial type \cite{MSK} may provide a very useful guidance.  Another interesting and important question is to understand the probability distribution of the value of the random field at the point of its global minimum. A few methods allowing to extract the large-deviation form for such a quantity based on the non-rigorous but powerful
 replica trick were proposed very recently in \cite{FLD2013} and shown to give reasonable results for the simplest case $p=2$. To extend that approach to the case $p>2$ and/or to the parabolically confined random landscapes remains an open issue.

\subsection*{Appendix: On the number of stationary points of an anisotropic stationary Gaussian random field.}

Introduce a postive-definite diagonal matrix $A=\mbox{diag}(a_1,\ldots,a_N)>0$ and consider Gaussian (mean zero) random field $V(\mathbf{x}), \,\, \mathbf{x}\in \mathbb{R}^N$  with the covariance structure $\mathbb{E}\left\{V(\mathbf{x}) V(\mathbf{y})\right\} = F\left(C(\mathbf{x},\mathbf{y})\right),$ where
\begin{equation}\label{1}
  C(\mathbf{x},\mathbf{y})=\frac{1}{2}(\mathbf{x}-\mathbf{y})^T\,A\,(\mathbf{x}-\mathbf{y})=\frac{1}{2} \sum_{i=1}^N a_i(x_i-y_i)^2
\end{equation}
and we will assume $F(C)$ being a smooth function suitably decaying at infinity.
Such a field is obviously stationary, but anisotropic.
We have
\begin{equation}\label{2}
 \mathbb{E}\left\{\frac{\partial}{\partial x_i} V(\mathbf{x}) V(\mathbf{y})\right\} =a_i(x_i-y_i)F'\left(C(\mathbf{x},\mathbf{y})\right)
\end{equation}
where the dash stands for the derivative over the argument. Differentiating once more gives
\begin{equation}\label{3}
  \mathbb{E}\left\{\frac{\partial}{\partial x_i} V(\mathbf{x}) \frac{\partial}{\partial y_j}V(\mathbf{y})\right \} =-a_i\delta_{ij}F'\left(C(\mathbf{x},\mathbf{y})\right)-a_ia_j(x_i-y_i)(x_j-y_j)F''\left(C(\mathbf{x},\mathbf{y})\right)
\end{equation}
and further
\begin{equation}\label{4}
 \mathbb{E}\left\{\frac{\partial^2}{\partial x_i\partial x_k} V(\mathbf{x}) \frac{\partial}{\partial y_j}V(\mathbf{y})\right\} =-a_ia_ja_k(x_i-y_i)(x_j-y_j)(x_k-y_k)F'''\left(C(\mathbf{x},\mathbf{y})\right)
\end{equation}
\[
-F''\left(C(\mathbf{x},\mathbf{y})\right)\left[a_ia_k\delta_{ij}(x_k-y_k)+a_ia_j\delta_{ik}(x_j-y_j)+
a_ia_j\delta_{jk}(x_i-y_i)\right]
\]
Now setting $\mathbf{x}=\mathbf{y}$ in the above we see that the vector of first derivatives $\tilde{v}_i=\frac{\partial}{\partial x_i} V(\mathbf{x})$ has diagonal covariance structure and, most importantly, is locally uncorrelated with the Hessian matrix with entries $\tilde{H}_{ik}=\frac{\partial^2}{\partial x_i\partial x_k} V(\mathbf{x})$:
\be\label{5}
 \mathbb{E}\left\{\tilde{v}_i\tilde{v}_j\right\} =-a_i\delta_{ij}F'(0), \quad  \mathbb{E}\left\{\tilde{H}_{ik}\tilde{v}_j\right\}=0
\ee
Finally, differentiating (\ref{4}) once again and setting $\mathbf{x}=\mathbf{y}$ we find the covariance structure
of the Hessian entries:
\begin{equation}\label{6}
 \mathbb{E}\left\{\tilde{H}_{ik}\tilde{H}_{jl}\right\}
 =F''\left(0\right)\left[a_ia_k\delta_{ij}\delta_{kl}+a_ia_j\delta_{ik}\delta_{jl}+
a_ia_j\delta_{jk}\delta_{il}\right]
\end{equation}
Now we make the following observation: consider the matrix $H$ defined in terms of the Hessian $\tilde{H}$ as $H=A^{-1/2}\tilde{H}A^{-1/2}$. We obviously have $H_{ij}=a_i^{-1/2}\tilde{H}_{ij}a_j^{-1/2}$ and a straightforward check shows that (\ref{6}) implies that the new matrix $H$ has the covariance structure:
\begin{equation}\label{7}
 \mathbb{E}\left\{H_{ik}H_{jl}\right\}  =F''\left(0\right)\left[\delta_{ij}\delta_{kl}+\delta_{ik}\delta_{jl}+\delta_{jk}\delta_{il}\right]
\end{equation}
 which is exactly the same as for the isotropic case $a_i=1, \forall i=1,\ldots, N$, see \cite{Fyo04}.

The mean number of stationary points of such field in any domain $D$ of the Euclidean space is given by
${\cal N}_s(D)=\int_D \rho_s({\bf x}) \, d{\bf x}$, where the density is given by the
Kac-Rice formula \ref{KR}. Using (\ref{5}) we easily get:
\be\label{8}
\mathbb{E}\left\{
\prod_{i=1}^N\delta\left(\frac{\partial}{\partial x_i}V\right)\right\}=\frac{1}{\left[2\pi |F'(0)|\right]^{N/2}}\frac{1}{\sqrt{\det{(A)}}}
\ee
whereas the use of (\ref{7}) implies:
\be\label{9}
 \mathbb{E}\left\{|\det{\left(\frac{\partial^2}{\partial x_i\partial x_k} V\right)}| \right\}=
\mathbb{E}\left\{|\det{\tilde{H}}| \right\}=\mathbb{E}\left\{|\det{A^{1/2}HA^{1/2}}| \right\}
\ee
\[
=\det{(A)}\cdot\mathbb{E}\left\{|\det{H}| \right\}
\]
Combining this two formulae we thus conclude that for any domain $D$ the mean number of stationary points for a stationary anisotropic field of such type is simply
proportional to the result for its isotropic counterpart:
\be\label{final}
{\cal N}^{(anis)}_s(D)=\sqrt{\det{(A)}}\, {\cal N}^{(iso)}_s(D)
\ee
where ${\cal N}^{(iso)}_s(D)$ for the isotropic case can be evaluated following the methods of \cite{Fyo04}. Similar proportionality obviously holds for stationary points of any given index, in particular for minima.

\end{document}